\documentclass[11pt,a4paper]{article}
\usepackage[T1]{fontenc}
\usepackage{lmodern}		
\usepackage[utf8]{inputenc}

\usepackage[a4paper,margin=1in]{geometry}

\usepackage{authblk}
\usepackage{amsmath}
\usepackage{amsfonts}
\usepackage{amsthm}
\usepackage{mathtools}

\usepackage{tocloft}

\usepackage{graphicx}
\usepackage{booktabs}

\usepackage[
    backend=biber,
    style=numeric-comp,
    sorting=nyt,
    giveninits=true,
    maxbibnames=99,
    doi=true,
    url=false,
    eprint=true
]{biblatex}

\DeclareBibliographyAlias{legislation}{misc}

\usepackage{xurl}
\usepackage{xcolor}
\usepackage[hidelinks]{hyperref}
\usepackage[capitalise,noabbrev]{cleveref}

\hypersetup{
    colorlinks=true,
    linkcolor=blue,
    citecolor=blue,
    urlcolor=blue
}

\numberwithin{equation}{section}

\usepackage{xcolor}

\usepackage{enumitem}

\setlist[itemize]{
    itemsep=0.1em,
    parsep=0pt,
    topsep=0.3em,
    partopsep=0pt
}

\newcommand{\iA}{\mathcal{A}}
\newcommand{\iF}{\mathcal{F}}
\newcommand{\iN}{\mathcal{N}}
\newcommand{\iT}{\mathcal{T}}

\newcommand{\iC}{\mathcal{C}}
\newcommand{\iL}{\mathcal{L}}
\newcommand{\dd}{\,\mathrm{d}}
\newcommand{\FF}{\mathbb{F}}
\newcommand{\EE}{\mathbb{E}}
\newcommand{\NN}{\mathbb{N}}

\newcommand{\RR}{\mathbb{R}}

\newcommand{\ZBO}{\textbf{ZBO}}

\newcommand{\IRS}{\textbf{IRS}}
\newcommand{\ESWO}{\textbf{ESWO}}
\newcommand{\BSWO}{\textbf{BSWO}}

\newcommand{\given}{\;\middle|\;}
\DeclarePairedDelimiterX{\E}[1]{\lbrack}{\rbrack}{\mathbb{E} #1}

\usepackage[bottom]{footmisc}
\usepackage{threeparttable}
\usepackage{makecell}

\usepackage{booktabs}
\usepackage{multirow}
\usepackage{caption}

\newcommand{\splus}{\textbf{$++$}}
\newcommand{\lplus}{\textbf{$+$}}
\newcommand{\neutral}{\textbf{$\circ$}}
\newcommand{\lminus}{\textbf{$-$}}
\newcommand{\sminus}{\textbf{$--$}}

\usepackage{algorithm}
\usepackage[indLines=true,noEnd=false]{algpseudocodex}

\usepackage{csquotes}

\usepackage{siunitx}
\DeclareSIUnit\bp{bp} 

\usepackage{epigraph}

\title{\textbf{Calibration of the Hull--White One-Factor Model for amortizing Bermudan swaptions: How much calibration is actually necessary?}}
\author{Kamil Domaga\l{}a%
\thanks{The views expressed in this paper are solely those of the author and do not necessarily reflect those of the affiliated institutions.}%
\hspace{0.4em}%
\thanks{Comments and suggestions are welcome at \href{mailto:domagala@htw-berlin.de}{domagala@htw-berlin.de}.}}
\affil{\vspace{1em}Martin Luther University Halle-Wittenberg, Halle (Saale), Germany}
\affil{HTW University of Applied Sciences, Berlin, Germany}
\affil{Investitionsbank Berlin, Berlin, Germany}
\date{November 26, 2026}
\begin{document}
%
%
%
\maketitle
%
%
%
\begin{abstract}
Researchers and practitioners have been using short-rate models to value Bermudan swaptions for many years, particularly because of their analytical tractability. For this purpose, parameter calibration is typically performed using the so-called core European swaptions underlying the Bermudan swaption. Since the market does not provide consistently available prices for amortizing structures, the approach is unsuitable for amortizing Bermudan swaptions without further adjustments. One way the literature addresses this gap is by generating synthetic target prices using basket replication. The approach relies on the repeated use of intermediate model instances, which have to be calibrated separately. The nested calibration runs make the valuation procedure computationally expensive. The contribution of this paper is twofold. First, we investigate a simple refinement of the basket replication method aimed at eliminating the nested calibration and therefore reducing the overall computational load. More specifically, the refinement replaces the resulting synthetic prices with upper-bound approximations. Second, using a multi-year historical dataset, we analyze to what extent different calibration sets and their cardinality materially affect the valuation of amortizing Bermudan swaptions under the Hull--White One-Factor model. Our findings suggest that reducing computational complexity, whether by restricting the cardinality of the calibration set or by relying on upper bounds as target prices during basket calibration, can lead to a substantially more efficient pricing procedure without materially affecting the corresponding valuations.
\end{abstract}
%
%
%
%
%
\medskip 

%
%
%
%
%
%
\section{Introduction}
\epigraph{
- When it comes to options there often isn't one correct price. \\
- True, but there surely are some obviously wrong ones!
}{
\textit{--- Conversation with a colleague, ca. 2018}}
\par The above exchange unfolded between two risk managers during a Bermudan swaption rollout. In our opinion, it accurately captures the idea that, in practice, option pricing tends to be more of a balancing act than a strict implementation of textbook algorithms. Interest rate models differ in their ability to incorporate market information relevant to pricing from a theoretical perspective. In general, a model's capability to utilize more information than other models seems to be a positive characteristic. But does the same hold for one individual model? Specifically, should we always prioritize completeness of the calibration set and the precision of target prices, often at the cost of increased complexity and reduced computational performance?
\par In this paper, we concentrate on parameter calibration of the Hull--White One-Factor (HW1F) model, using a PDE-based approach for pricing. The discussion of whether a short-rate model can be used to appropriately value a Bermudan option is beyond the scope of this paper. At the same time, we acknowledge that this question has been occupying researchers and practitioners for many years. For an overview of arguments on both sides, we refer to the work of Gatarek and Jab\l{}ecki~\cite{gatarekScyllaCharybdisBermudan2021}. Among other things, the work wittily places the famous debate on the suitability of one-factor models for Bermudan swaption pricing between Andersen and Andreasen~\cite{andersen2001FactorDependenceBermudan} and Longstaff et al.~\cite{longstaff2001ThrowingAwayBillion} in the context of Greek mythology.
\par For short-rate models, parameter calibration is of particular importance since it is the most effective way to reflect relevant market information.  In the context of Bermudan swaptions, the underlying European swaptions are natural candidates for calibration. Since the market does not provide consistently available prices for amortizing structures, this strategy is unsuitable for amortizing Bermudan swaptions without further adjustments. One way the literature addresses this gap is by generating synthetic target prices using basket replication (see, e.g., \cite[pp.~887--889]{andersen2010InterestRateModeling} and \cite[pp.~46--49]{haganMETHODOLOGYCALLABLESWAPS2004}). The approach relies on the repeated use of intermediate model instances, which have to be calibrated separately. The nested calibration runs make the valuation procedure computationally expensive. The computational costs often limit the practical applicability of the approach. 
\par The literature on the Hull--White model is extensive and surprisingly scarce at the same time. The theoretical properties and the valuation procedures are thoroughly covered in many textbooks and articles focusing on specific aspects of the model. However, existing research provides limited empirical evidence comparing pricing results based on different calibration sets. While Gurrieri et al.~\cite{gurrieri2009CalibrationMethodsHullWhite} present some numerical results for different currencies and parametrizations, their work focuses primarily on the fitted parameters and the ability of the model to reconstruct the target prices rather than on Bermudan swaption prices corresponding to various calibration sets. Albeit on a limited scale, numerical results comparing different calibration approaches have been reported in practitioner-oriented publications, such as Steinberg and Wolff-Siemssen~\cite{steinberg2024InterestRateOptionPricing}.  
\par With this in mind, the purpose of this paper is twofold. First, we investigate a simple refinement of the basket replication method aimed at eliminating the nested calibration runs and therefore reducing the overall computational load of the approach. More specifically, following the practitioner presentations by Miemiec~\cite{miemiec2013accretingSwaptions} and Kürzinger~\cite{kuerzinger2017aspectsIrregularSwaptions}, we analyze whether the replacement of synthetic prices by an upper-bound approximation yields consistent valuations across a broad range of products and market data scenarios. To the best of our knowledge, the suitability of this approximation as a calibration target has not been systematically investigated in either presentation. Roos~\cite{roos2016ModelIndependentBounds} examined a related approximation approach applied directly to amortizing Bermudan swaptions using a basket of bullet Bermudan swaptions. The study reported promising results indicating that the approximations are tight enough for meaningful arbitrage testing and potentially even for static hedging strategies. Second, using a historical dataset of EUR interest-rate market data, we systematically analyze to what extent different calibration sets and their cardinality materially affect the valuation of amortizing Bermudan swaptions under the Hull--White One-Factor model. To this end, we consider thousands of amortizing Bermudan swaptions randomly generated across a broad spectrum of realistic contract characteristics. Moreover, the products are valued under randomly assigned market data scenarios observed between December 2021 and November 2025.
\par Our findings suggest that reducing computational complexity, whether by restricting the cardinality of the calibration set or by relying on upper bounds as target prices during basket calibration, can lead to a substantially more efficient pricing procedure without materially affecting the corresponding valuations. Furthermore, we find that notional-sensitive calibration sets exhibit the highest level of agreement, indicating consensus on the overall price level for amortizing swaptions.
\par The remainder of the paper is organized as follows. In Section~\ref{sec:methodology}, we outline the HW1F model specification, formally introduce the Bermudan swaption, and define the examined calibration sets. In Section~\ref{sec:empirical_results}, the experimental setting is described and the empirical results are discussed. Finally, Section~\ref{sec:conclusions} contains the main conclusions. 
%
%
%
%
%
\section{Methodology}\label{sec:methodology}
\subsection{Hull--White One-Factor Model}\label{sec:hw1f_model}
\par A large number of textbooks and academic papers provide a detailed treatment of the Hull--White model. In this subsection, we do not intend to recapitulate this long-standing literature but rather outline the model with the goal of establishing the nomenclature, introducing the functional form of the parameters, and indicating the implementation strategy. To make the paper self-contained, we include all derivations related to the HW1F model in Appendix~\ref{app:hw1f_formulas}. For further details, we direct the reader to the standard references \cite[pp.~71--80]{brigo2006InterestRateModels}, \cite[pp.~364--387]{bjork2009ArbitrageTheoryContinuous}, and \cite[pp.~415--431]{andersen2010InterestRateModeling}, which range from a general introduction to a more practitioner-oriented framework.  
\par The exact specification of the short rate evolution $r(t)$ of the HW1F model has varied over the years since its introduction in \cite[pp.~573--592]{hull1990PricingInterestRateDerivativeSecurities}. This paper adopts the following formulation, also known as the extended Vasicek model in the literature:  
\begin{align}\label{eq:hw1f_dynamic}
\dd r(t) & = \left[ \vartheta (t) - a(t)r(t) \right]\dd t+\sigma (t) \dd W (t) \nonumber \\ 
\text{with}\quad r(0) & = r_0,
\end{align}
where the functions $\vartheta (t)$, $a(t)$, and $\sigma (t)$, referred to as drift rate, mean reversion speed and diffusion coefficient, respectively, are deterministic functions of time. Moreover, the process $(W(t))_{t\ge 0}$ denotes a Brownian motion on the filtered probability space $(\Omega, \iF, \FF, Q)$, where $\FF =(\iF_t)_{t\ge 0}$ satisfies the usual conditions and $Q$ denotes the risk-neutral measure associated with the bank account acting as the numéraire.
\par Building on these definitions, we now turn to the precise specification of the parameters. The drift rate $\vartheta (t)$ is set to replicate the currently (at $t=0$) observed interest rate structure\footnote{Introducing the additional function $\vartheta(t)$ into the drift term of the differential equation is known as \textit{deterministic shift extension}.} and is equal to   
\begin{align}\label{eq:hw1f_drift_term_result}
    \vartheta(t) & = a(t)f^\mathrm{M}(0,t) + \frac{\partial}{\partial t} f^\mathrm{M}(0,t) + V(0,t) \nonumber \\
    \text{with}\quad V(t,T) & \coloneqq \int_t^T \sigma^2 (u) e^{-2\iA (u,T)}  \dd u,
\end{align}
where $\iA (t,T) = \int_t^Ta(u)\dd u$ and $f^\mathrm{M}(t, T)$ denotes the market value of the instantaneous forward rate for maturity $T>t$ observed at time $t$. Note that in this formulation the short rate process $r(t)$ reverts to the time-dependent term $\frac{\vartheta(t)}{a(t)}$. This leaves us with only $a(t)$ and $\sigma(t)$ as free parameters. We allow both functions to be piecewise constant, i.e. 
\begin{equation*}
\begin{aligned}a(t) & \coloneqq \sum_{i=1}^{n+1} a_i \mathbf{1}_{ [ t^a_{i-1}, t^a_{i}) }(t) \quad\text{and}\quad \sigma (t) & \coloneqq  \sum_{i=1}^{m+1} \sigma_i \mathbf{1}_{ [ t^{\sigma}_{i-1}, t^{\sigma}_{i}) }(t)
\end{aligned}
\end{equation*}
with intervals defined by $0=t_0^a<t_1^a<\cdots  <t_{n+1}^a=\infty$ and $0=t_0^\sigma<t_1^\sigma <\cdots<t_{m+1}^\sigma = \infty$ as well as constant values $a_1,\ldots, a_{n+1}$ and $\sigma_1,\ldots, \sigma_{m+1}$, satisfying $a_{n+1} = a_n$ and $\sigma_{m+1} = \sigma_m$, respectively. The step functions are defined so that $n$ and $m$ correspond to the respective numbers of instruments used in the calibration, while flat extrapolation is imposed beyond the last step. A detailed derivation of all values will be discussed in Section~\ref{sec:calibration_framework}. 
\par Although we allow both the mean reversion speed $a(t)$ and the diffusion coefficient $\sigma (t)$ to be step functions, the pricing relies on model instances with only one of the parameters being piecewise constant while the other is strictly constant. This is motivated by the observation that time-dependent parameters, in general, tend to produce forward rate volatilities that do not necessarily agree with empirical evidence. Therefore, the literature usually recommends setting the mean reversion speed to a constant value $a(t) \equiv a $ and calibrating the diffusion coefficient to market observations, which supports a consistent shape of the forward rate volatilities. This will be our standard model parametrization for the pricing procedure of Bermudan swaptions. As part of the calibration set construction, we will leverage intermediate model instances with piecewise constant mean reversion speed and a constant diffusion coefficient to obtain target prices for amortizing instruments. For a broader discussion of this matter, we refer to \cite[pp.~418--420]{andersen2010InterestRateModeling}.
\par One final remark is in order. Introducing the HW1F model usually involves solving the differential equation given by \eqref{eq:hw1f_dynamic}, deriving model prices $P(\cdot,\cdot)$ of zero-coupon bonds, and setting the drift term to match market observations $P^\mathrm{M}(\cdot, \cdot)$, i.e. $P(0,T) \overset{!}{=} P^\mathrm{M}(0, T)$ for all $T>0$. This is usually demonstrated for continuous functions $a(t)$, $\sigma(t)$, $\vartheta(t)$ and implicitly $f^\mathrm{M}(0,t)$. The general results remain valid for piecewise constant (and therefore not everywhere continuous) parameters. From a numerical perspective, the evaluation of terms like $\frac{\partial}{\partial t}f^\mathrm{M}(0, t)$ in \eqref{eq:hw1f_drift_term_result} is problematic. This is due to the fact that the differentiability of $f^\mathrm{M}(0, t)$ depends on the construction method of the yield curve. To overcome this problem, the implementation is based on an auxiliary process given by $x(t) \coloneqq r(t)-f^\mathrm{M}(0,t)$ with $x(0) = 0$. By It\^{o}'s Lemma, we get the following time evolution
\begin{equation}\label{eq:auxiliary_state_variable_hw1f}
\dd x(t) = \left[ V(0,t) - a(t)x(t) \right]\dd t + \sigma( t) \dd W(t).  
\end{equation}
This eliminates the necessity to evaluate the problematic term. The resulting closed-form expression for the price of a zero-coupon bond $P(t,T)=P(t,T;x(t))$ is given by
\begin{equation}
\begin{split}
    P(t,T;x(t)) &= \frac{P^\mathrm{M}(0,T)}{P^\mathrm{M}(0,t)} \exp \left( - x(t)B(t,T) - \frac{1}{2}B^2 (t,T)V(0,t) \right) \\
    \text{with } B(t,T) &= \int_t^T e^{-\iA(t,u)}\dd u.
\end{split}
\label{eq:hw1f_zero_coupon_bond_formula}
\end{equation}
Note that the term $P(t,T;x(t))$ describes the price of a zero-coupon bond at a future time $t>0$ given the value of the state variable $x(t)$ at that time. In particular, it should not be confused with $P^\mathrm{M}(0,T)$ observed in the market at $t=0$. For completeness, a detailed derivation of the preceding results is included in Appendix~\ref{app:hw1f_formulas}.
\subsection{Bermudan Swaption}\label{sec:bswo}
\par As a preliminary step, we briefly address cash flow generation in interest rate derivatives and introduce an interest rate swap before turning to a formal definition of a swaption. 
\par \textbf{Schedule, Conventions and IBOR-type Rates}\newline A schedule $\iT=\{T_i\}_{i=0}^n$ over $[T_0, T_n]$ is a series of dates spanning from $T_0$ to $T_n$, i.e. $T_0 <T_1<\cdots <T_n$. In the context of interest rate derivatives, the dates are usually generated by rolling a tenor backwards from the maturity date and subsequently adjusting the dates according to the business day convention specified by the contract.\footnote{Note that the business day convention is applied after all the dates have been generated. By construction, the unadjusted dates are equally spaced except possibly the first period often referred to as a stub.} Given a time point $T\in [T_0, T_n)$ and a schedule $\iT=\{T_i\}_{i=0}^n$, we define the sequence $\iT\vert_T$ consisting of $T$ followed by all subsequent dates from $\iT$, i.e. $T,T_m,T_{m+1},\ldots,T_n$ with $m = \min\{i: T_i>T\}$, as the $T$-truncated schedule. The time span between two dates of a schedule is expressed using year fractions. In particular, we set $\tau_i\coloneqq \tau (T_{i-1}, T_i)$ for $i=1, \ldots, n$ with $\tau (\cdot, \cdot)$ specified by the underlying day count convention. We refer the reader to \cite[pp.~15--22, 89--91]{henrard2025InterestRateInstruments} for an overview of market conventions in the context of interest rate derivatives. 
\par A simply compounded spot IBOR-type rate over $[T_1,T_2]$ is given by the expression
\begin{equation*}
    L (T_1,T_2)\coloneqq \frac{1-P(T_1,T_2)}{\tau(T_1,T_2)P (T_1,T_2)}.    
\end{equation*}
Analogously, a simply compounded forward IBOR-type rate over $[T_1,T_2]$ observed at $t<T_1$ is given by 
\begin{equation*}
    L (t,T_1,T_2) \coloneqq \frac{P(t,T_1)-P(t,T_2)}{\tau(T_1,T_2)P (t,T_2)}.    
\end{equation*}
The above rates serve as a mathematical representation of reference interest rate indices used to determine floating-rate payments of many financial instruments. This applies in particular to the EURIBOR indices widely employed in euro-denominated contracts considered in this paper. Given $t<T_1<T_2$, the spot rate $L(T_1,T_2)$ represents the future random rate that is fixed at $T_1$.\footnote{To ease the notation, we ignore the fixing lag.} The forward rate $L(t,T_1,T_2)$ denotes the rate implied by the term structure observed at $t$.   
\par To account for a multi-curve setting, we will include an additional label on interest rate objects, e.g., $L^\Delta (T_1,T_2)$ or $P_\mathrm{c}$. The super- or subscript identifies the curve associated with the object. The $\Delta$ indicates the curve corresponding to the calendar tenor used to generate the underlying schedule, i.e. $\Delta \approx T_2-T_1$. The $c$ stands for the curve consistent with the collateral agreement of the contract. For the purposes of this study, we restrict $\Delta$ to 3 or 6 months and use the €STR-based curve for discounting. For an in-depth discussion of multi-curve bootstrapping,~see~\cite{ametranoEverythingYouAlways2013}.
\par \textbf{Interest Rate Swap}\newline An interest rate swap $\IRS$ is an agreement between two parties to exchange interest-rate payments based on two cash flow legs (fixed and floating) over a predefined time period $[T^\mathrm{S}, T^\mathrm{M}]$. Each leg is specified in terms of an individual schedule over $[T^\mathrm{S},T^\mathrm{M}]$ as follows:
\begin{itemize}
    \item Fixed leg: consists of interest rate payments $N_j\tau_j S$ calculated over the accrual period $[T_{j-1}, T_j]$ paid at $T_j$ for $j=1,2, \ldots, n$ with tenor schedule $\iT=\{T_j\}_{j=0}^n$, notional schedule $\iN = \{ N_1, N_2, \ldots,$ $N_n  \} $, fixed rate $S$, and year fractions $\tau_j= \tau (T_{j-1}, T_j)$.
    \item Floating leg: consists of interest rate payments $\tilde{N}_i\tilde{\tau}_i L^\Delta (\tilde{T}_{i-1}, \tilde{T}_i)$ calculated over the accrual period $[\tilde{T}_{i-1}, \tilde{T}_i]$ paid at $\tilde{T}_i$ for $i=1,2, \ldots, m$ with tenor schedule $\tilde{\iT}=\{\tilde{T}_i\}_{i=0}^m$, notional schedule $\tilde{\iN} = \{ \tilde{N}_1, \tilde{N}_2, \ldots,  \tilde{N}_m  \} $, reference index $L^\Delta(\cdot, \cdot)$ and year fractions $\tilde{\tau}_i= \tilde{\tau} (\tilde{T}_{i-1},\tilde{T}_i)$. 
\end{itemize}
\par A few remarks on the above definition are in order. If a party pays the fixed rate $S$, it has entered a payer swap. Conversely, if the party receives fixed payments, the swap is called a receiver swap. Although not directly implied by the above specification, we restrict the notional schedules $\iN$ and $\tilde\iN$ to result in identical notionals for any time point in $[T^\mathrm{S}, T^\mathrm{M}]$. This implies that the notional can only change on dates that exist in both schedules simultaneously. An interest rate swap constructed on a set of standard conventions, without any optionality and in particular with constant notional is referred to as plain vanilla. 
\par The $t<T^\mathrm{S}$ present value of an interest rate swap $\IRS_t$ is given by the following expression
\begin{equation*}
\IRS_t ( \iT, \tilde{\iT}, S, L^\Delta,\iN, \tilde{\iN}, \tau, \tilde{\tau}, \omega) = \omega \left[ \sum_{j=1}^n N_j \tau_j SP_\mathrm{c}(t,T_j)- \sum_{i=1}^m \tilde{N}_i L^{\Delta}(t, \tilde{T}_{i-1}, \tilde{T}_i) \tilde{\tau}_i P_\mathrm{c}(t,\tilde{T}_i) \right]
\end{equation*}
with $\omega = +1$ for a receiver swap and $\omega = -1$ for a payer swap. In practice, it is often convenient to price swap contracts in terms of the so-called par rate $S^\mathrm{p}(t)\coloneqq S^\mathrm{p}(t, \iT, \tilde{\iT}, L^\Delta,\iN, \tilde{\iN}, \tau, \tilde{\tau})$ for which the swap has a zero value. This allows us to simplify the above pricing formula into 
\begin{align}\label{eq:irs_price_through_swap_rate}
    \IRS_t ( \iT, \tilde{\iT}, S, L^\Delta,\iN, \tilde{\iN}, \tau, \tilde{\tau}, \omega) & = \omega \left[ S-S^\mathrm{p}(t)\right] \sum_{j=1}^n N_j\tau_jP_\mathrm{c}(t,T_j),
\end{align}
where we refer to the difference $M(t) \coloneqq S-S^\mathrm{p}(t)$ as moneyness and $C(t) \coloneqq \sum_{j=1}^n N_j\tau_jP_\mathrm{c}(t,T_j)$ as annuity. To ease the notation, we will later drop some of the input parameters not essential for the problem at hand. Particularly for calibration, it will be convenient to reduce the contract specification to the period $[T^\mathrm{S}, T^\mathrm{M}]$, moneyness $M$ and constant notional $N$ resulting in $\IRS_t(T^\mathrm{S},T^\mathrm{M},M, N)$. The omitted parameters will be outlined before the reduced notation is used.
\par \textbf{Swaption Definition}\newline A swaption is an agreement between two parties that gives the holder (buyer) the right, but not the obligation, to enter into a predefined interest rate swap $\IRS( \iT, \tilde{\iT}, S, L^\Delta,\iN, \tilde{\iN}, \tau, \tilde{\tau}, \omega)$ at a predefined date (settlement date) without any additional costs upon exercise. A swaption is called European $\ESWO$ if the holder can enter into the swap at a single point in time $T^\mathrm{S}$ and Bermudan $\BSWO$ if the holder can choose one point from a settlement schedule $\{T^\mathrm{S}_k\}^p_{k=1}\subseteq[T^\mathrm{S},T^\mathrm{M})$. The exercise decision has to be declared irrevocably at the exercise date $T^\mathrm{E}\le T^\mathrm{S}$ or $T^\mathrm{E}_k\le T^\mathrm{S}_k$ for $k=1,\ldots,p$, respectively.
\par Before discussing the valuation procedure of swaptions used in this paper, a few comments on the product definition are in order. First, a Bermudan swaption can be interpreted as a collection of European swaptions from which only one can be exercised.\footnote{The possibility of a partial exercise is ignored intentionally.} If the holder fails to exercise, the option expires worthless. Throughout the remainder of this paper, the European swaptions from this collection are referred to as core swaptions of the Bermudan swaption. Next, analogously to the cash flow schedule, the settlement dates $\{T^\mathrm{S}_k\}^p_{k=1}$ usually result from a calendar tenor (exercise frequency) rolled backwards from the maturity of the underlying swap. The corresponding exercise dates $\{T^\mathrm{E}_k\}^p_{k=1}$ are obtained through shifting the settlement dates by the so-called notice period. 
\par Lastly, since we did not require $\{T^\mathrm{S}_k\}^p_{k=1}\subseteq\iT\cap\tilde{\iT}$, the settlement can occur in the middle of an interest rate period, raising questions about the precise specification of the exercised-into underlying swap. In this paper, we adopt the assumption that the holder enters into a swap resulting from the truncated schedules $\iT\vert_{T^\mathrm{S}_k}$ and $\tilde{\iT}\vert_{T^\mathrm{S}_k}$. With all other conventions unchanged, at $T_k^\mathrm{S}$ the holder formally settles into
\begin{equation*}
    \IRS( \iT\vert_{T^\mathrm{S}_k}, \tilde{\iT}\vert_{T^\mathrm{S}_k}, S, L^\Delta,\iN\vert_{T^\mathrm{S}_k}, \tilde{\iN}\vert_{T^\mathrm{S}_k}, \tau, \tilde{\tau}, \omega)\quad\text{for}\quad k=1, \ldots, p 
\end{equation*}
with the notional schedules truncated analogously. Note that we consider the resulting swaps as fully specified stand-alone instruments. The potentially resulting stubs are therefore treated as ordinary self-contained accrual periods. This means that the forward rates will be projected from the same curve based on their actual length. In particular, the original fixing or year fractions are not inherited.\footnote{To the best of our knowledge, at least two other mechanisms exist in practice. To begin with, such settings are avoided by design. The schedules are either identical or coincide on the relevant dates. This can be achieved by alignment of the respective rolling periods and business day conventions. The second approach requires the exchange of all accrued interest at the settlement date. The holder then settles into the remaining cash flows of the unique underlying swap. For further details on this, we refer to \cite[pp.~5--9]{haganMETHODOLOGYCALLABLESWAPS2004}.} 
\par \textbf{European Swaption Valuation}\newline To derive the HW1F price of a European swaption $\ESWO_t (T^\mathrm{E}, T^\mathrm{S}, \IRS)$ at time $t<T^\mathrm{E}$, we first isolate the basis spread between the discounting and forwarding curves entering the payoff. To this end, a valuation-date-dependent rate $S^\mathrm{BS}(t)\coloneqq S^\mathrm{BS}(t, \iT, \tilde{\iT}, L^\Delta,\iN, \tilde{\iN}, \tau, \tilde{\tau})$ is calculated such that 
\begin{equation*}
\IRS_t (\iT, \tilde{\iT}, S, L^\Delta,\iN, \tilde{\iN}, \tau, \tilde{\tau}, \omega)  \overset{!}{=} \IRS_t (\iT, \tilde{\iT}, S+S^\mathrm{BS}(t), L^\mathrm{c},\iN, \tilde{\iN}, \tau, \tilde{\tau}, \omega). 
\end{equation*}
\par Subsequently, we assume that in the context of option pricing the basis spread does not evolve over time, i.e. $S^\mathrm{BS}(t)\equiv S^\mathrm{BS}$. With this simplification, we can rearrange the terms in the payoff to represent a coupon bond
\begin{align}\label{eq:pv_swap_as_bond}
    \IRS_t & = \omega \left[ \sum_{j=1}^n N_j \tau_j (S+S^\mathrm{BS} )P_\mathrm{c}(t,T_j)+ \sum_{i=0}^m \left( \tilde{N}_i - \tilde{N}_{i+1}\right) P_\mathrm{c}(t,\tilde{T}_i) \right] \nonumber\\ 
    &  = \omega \sum_{j=0}^n \left[ N_j \left(  1+\tau_j\left( S+S^\mathrm{BS} \right) \right) -N_{j+1}\right] P_\mathrm{c}(t,T_j) \nonumber \\
    & \eqqcolon \omega \sum_{j=0}^n c_jP_\mathrm{c}(t,T_j)
\end{align}
with the convention $\tilde{N}_0=\tilde{N}_{m+1} = N_0=N_{n+1}=\tau_0=0$ and coupons $c_j$ for $j=0,\ldots,n$. The second equality holds due to the previously stated restriction that notional adjustments must be applied simultaneously to both legs. Note that a payer or receiver swap is equivalent to a short or long coupon bond position, respectively.
\par Finally, applying Jamshidian's decomposition\footnote{The approach relies on the existence of a unique exercise boundary $x^*$. The literature usually states two jointly sufficient conditions for this to be fulfilled: monotonicity of $P(t,T;x)$ in $x$ combined with strictly positive coupon payments. The setting of this paper violates the latter in two cases. First, we choose to include the first notional, usually incorporated in the strike, in the cash flows, resulting in a negative first coupon. This gives us the flexibility to reflect different exercise and settlement dates. However, its corresponding discount factor covers only the notice period, which limits its potentially disruptive effect on monotonicity. Second, if the swaption has a negative fixed rate that is not compensated by the amortization rate, individual coupons can become slightly negative. In the numerical experiments, a solution $x^*$ satisfying the exercise boundary condition was successfully identified in every valuation. Although uniqueness was not systematically investigated, no numerical irregularities suggesting a violation were observed.} (see \cite[pp.~205--209]{jamshidian1989ExactBondOption}) yields the present value of the European swaption as a weighted sum of zero-coupon bond options
\begin{align*}
    \ESWO_t(T^\mathrm{E},T^\mathrm{S},\IRS) & =  P_\mathrm{c}(t,T^\mathrm{E})\EE^{T^\mathrm{E}} \left[ \left(\omega \sum_{j=0}^n c_jP_\mathrm{c}(T^\mathrm{E},T_j) \right)^+ \given \iF_t  \right] \\
    & = \sum_{j=0}^n c_j \ZBO_t \left( T^\mathrm{E},T_j, P_\mathrm{c}(T^\mathrm{E},T_j,x^*), \omega \right),
\end{align*}
where $x^*\in\RR$ is chosen such that $0 = \sum_{j=0}^n c_j P_\mathrm{c}(T^\mathrm{E},T_j,x^*)$. The expectation $\EE^{T^\mathrm{E}}[\cdot]$ is taken under the $T^\mathrm{E}$-forward measure associated with $P_\mathrm{c}(\cdot, T^\mathrm{E})$ acting as numéraire. The price of the zero-coupon bond option under the HW1F model is given by 
\begin{equation}
\begin{split}
    \ZBO_t (T^\mathrm{E}, T^\mathrm{M}, K,\omega) &= \omega P(t,T^\mathrm{M}) \Phi \left( \omega \left( d + \sqrt{\theta} \right) \right) - \omega K P(t,T^\mathrm{E}) \Phi (\omega d) \\
    \text{with } d &= \frac{\ln\left(\frac{P(t,T^\mathrm{M})}{P(t,T^\mathrm{E})K} \right) - \frac{1}{2} \theta }{\sqrt{\theta}}, \\
    \theta &= B^2(T^\mathrm{E},T^\mathrm{M}) V(t,T^\mathrm{E}).
\end{split}
\label{eq:zbo_pricing_formula}
\end{equation}
For completeness, a formal definition of a $\ZBO$ contract and a detailed derivation of its pricing formula are included in Appendix~\ref{app:hw1f_formulas}.
\par \textbf{Bermudan Swaption Valuation}\newline Since a Bermudan swaption can be exercised on any of a specified set of dates, we now consider a series of payoffs $H_k$ at time points $T^\mathrm{E}_k$ given by 
\begin{align*}
    H_k & = \left[  \IRS_{T^\mathrm{E}_k}^k\right]^+ \\
     \text{with}\quad\IRS^k & \coloneqq \IRS( \iT\vert_{T^\mathrm{S}_k}, \tilde{\iT}\vert_{T^\mathrm{S}_k}, S, L^\Delta,\iN\vert_{T^\mathrm{S}_k}, \tilde{\iN}\vert_{T^\mathrm{S}_k}, \tau, \tilde{\tau}, \omega) \quad\text{for}\quad k= 1, \ldots, p.
\end{align*}%
If the right is exercised at the last possible date $T^\mathrm{E}_p$, the holder simply receives the immediate value of $H_p$. At any exercise date $T_k^\mathrm{E} < T_p^\mathrm{E}$, the holder has to decide between exercising and receiving $H_k$ immediately or continuing and postponing the exercise decision to any of the subsequent dates. Consequently, the exercise decision is dictated by comparison of $H_k$ and the value of the outstanding rights. From the perspective of the martingale pricing framework, we are facing an optimal stopping problem. The price of the Bermudan swaption $\BSWO_t ( \iT^\mathrm{E}, \iT^\mathrm{S},\IRS)$ at a time $t$ is then given by
\begin{align*}
\BSWO_t ( \iT^\mathrm{E}, \iT^\mathrm{S},\IRS) \coloneqq \sup_{ K \in \iT_t } \EE^Q \left[ \exp\left(  -\int_t^{T_K^\mathrm{E}} r(u) \dd u  \right) H_K \given \iF_t\right],
\end{align*}
where $\iT_t$ denotes the set of all admissible stopping time indices $K$ taking values in $\{ k\in\{1,\ldots,p\}:T_k^\mathrm{E}>t\}$. The solution to this type of problem is given by the Snell envelope, which defines a value process $(V_k)_{k=0}^p$ corresponding to $\{T^\mathrm{E}_k\}_{k=0}^p$ with $T^\mathrm{E}_0=t$ and $H_0=0$ set for convenience. Given $V_p =H_p$, the process is defined recursively by
\begin{align*}
V_k & \coloneqq \max ( H_k , C_k )\\
\text{with}\quad C_k & \coloneqq \EE^Q \left[ \exp\left(  -\int_{T^\mathrm{E}_k}^{T^\mathrm{E}_{k+1}} r(u) \dd u  \right) V_{k+1} \given \iF_{T_k^\mathrm{E}}  \right] \quad \text{for}\quad k = p-1, \ldots, 0. 
\end{align*}
The value $C_k$ is referred to as the continuation value. It reflects the value of postponing the exercise decision. At each step of the value process $(V_k)_{k=0}^p$ we compare the immediate exercise value $H_k$ with the continuation value $C_k$, as outlined above.
\par In practical applications, the process $(V_k)_{k=0}^p$ is usually approximated using one of the following three methods: Monte Carlo simulation, recombining trees, or finite difference solvers. The last of these will be used in this paper. The approach is based on the observation that the value of a European contingent claim, expressed as a conditional expectation in the martingale approach, can also be derived from a partial differential equation (PDE) with the payoff as the terminal condition. This connection is formally established by the Feynman--Kac theorem. In our particular case, given the state variable $x$ defined in \eqref{eq:auxiliary_state_variable_hw1f} and a contingent claim with terminal payoff $H(x)$ at $T$, the value function $u(t,x)$ of the European-style option is given by the solution to
\begin{align*}
    \frac{\partial}{\partial t}u(t,x) + \left[ V(0,t)-a(t)x\right]\frac{\partial}{\partial x}u(t,x)+\frac{\sigma^2(t)}{2}\frac{\partial^2}{\partial x^2} u(t,x)-\left(x+f^\mathrm{M}(0,t)\right)u(t,x) & = 0 \\
    \text{with}\quad u(T,x) & = H(x).
\end{align*}
This type of equation can be solved numerically using a finite difference solver. The basic idea is to discretize the equation on a predefined spatial and temporal grid, approximate the partial derivatives of the function by finite differences, and finally solve the resulting linear system of equations. Note that the terminal condition $H(x)$ requires the evaluation of the underlying swaps at the respective spatial grid points. This can be accomplished using the zero-coupon bond formula from \eqref{eq:hw1f_zero_coupon_bond_formula} to discount the cash flows obtained from the bond representation of the underlying swaps introduced in \eqref{eq:pv_swap_as_bond}.
\par The general procedure to approximate the value process of a Bermudan option $(V_k)_{k=0}^p$ can then be outlined as follows
\begin{itemize}
\item At the last exercise date $T_p^\mathrm{E}$ the price of the Bermudan option matches the  payoff $V_p=H_p$
\item For any date $t\in [T^\mathrm{E}_{p-1},T^\mathrm{E}_p)$ we can evaluate the European-style contingent claim  with payoff $H_p$ at $T_p^\mathrm{E}$. Note that for $t=T_{p-1}^\mathrm{E}$ this value matches exactly $C_{p-1}$. Since there are no exercise opportunities on $(T^\mathrm{E}_{p-1},T^\mathrm{E}_p)$ the prices of the Bermudan and European options coincide 
\item For $k=p-1,\ldots,1$, we repeat the following steps
\begin{itemize}
\item At $T^\mathrm{E}_{k}$ the price of the Bermudan option results from the comparison of the payoff and the continuation value, i.e. $V_k= \max ( H_k , C_k )$
\item For any date $t\in [T^\mathrm{E}_{k-1},T^\mathrm{E}_{k})$ we evaluate a European-style contingent claim with terminal payoff $\max ( H_k , C_k )$ at maturity $T^\mathrm{E}_{k}$. Note that for $t=T_{k-1}^\mathrm{E}$ this value matches exactly $C_{k-1}$
\end{itemize}
\item After the final step, the value $V_0 =\max(H_0,C_0)$ coincides with the previously defined price of the Bermudan option $\BSWO_t$
\end{itemize}
\par We conclude with a few remarks on the specific finite difference solver used in the numerical experiments below. The approach adopted is the Crank--Nicolson method, which utilizes central differences in the state dimension and is semi-implicit in the time dimension. The implementation follows a standard textbook description. Since the precise numerical protocol lies outside the scope of this paper, at this stage we only summarize the key aspects of the specification and refer to Appendix~\ref{app:crank_nicolson_method} for further details. We discretize the spatial variable $x$ on the interval $\qtyrange{-20}{20}{\percent}$ divided into 300 subintervals of equal size. The time dimension between a valuation date $t$ and the relevant exercise date $T^\mathrm{E}$ is partitioned using a step size of two calendar days, possibly resulting in an irregular first interval. These values were adopted following \cite[pp.~8--11]{leeFiniteDifferenceMethod2020}, where their stability was demonstrated. At the boundaries, we set the second derivatives to zero and approximate the first derivatives by forward or backward differences at the left or right boundary, respectively.
\subsection{Calibration Framework}
\label{sec:calibration_framework}
\par Having established the product specification, we build on the HW1F model introduced in Section~\ref{sec:hw1f_model} and turn to its calibration. The procedure is divided into two consecutive steps: construction of the calibration set and the numerical optimization of parameter values. Both steps have to be handled carefully to ensure reliable results. Typical weaknesses include prioritizing the number of calibration instruments over relevance and added value, ignoring market conventions, and uncritically accepting the fitted values. We discuss both steps in what follows, starting with the construction of the calibration set.
\subsubsection{Calibration-set Construction}
\par A calibration set $\iC$ is a set of $n\in\NN$ ordered pairs of instruments and target prices observed or derived from the market on a valuation date $t$, i.e. 
\begin{equation*}
    \iC \coloneqq \left\{ \left( I_1, \mathrm{PV}^\mathrm{M}_1\right),  \ldots , \left( I_n, \mathrm{PV}^\mathrm{M}_n\right)  \right\}.
\end{equation*}
In this paper, we only consider European swaptions as calibration instruments. Since the Bermudan swaption is the primary product class of this study, the closely related European swaptions act as a natural building block of the calibration set. The calibration sets considered henceforth fall into two high-level categories: local and global. The distinction is based on whether the choice of instruments in the calibration set depends on the valued product. Global methods calibrate the model parameters once based on a predefined basket of instruments and price all products with the same set of parameter values. By contrast, local approaches identify the parameter set based on a highly specific basket of instruments picked for each product individually. From a purely practical perspective, global approaches have the advantage of reduced computational effort because the numerical calibration has to be performed only once. However, for short-rate models product-specific calibration sets provide a straightforward way to ensure that market information relevant to that product is sufficiently reflected. Before discussing the different ways of constructing a calibration set, we briefly turn to the question of how prices of European swaptions are quoted in the market. 
\par \textbf{Volatility Cube}\newline Standardized market quotes exist only for a restricted set of European swaptions. First, quotes are available on a predefined discrete grid of exercise dates, underlying swap tenors and strike rates. Next, given scarce liquidity, the market supplies quotes only for options on swaps with a constant notional schedule. Moreover, the options and the underlying swaps follow a strict set of conventions, e.g., spot period, floating index, payment frequencies or day count conventions. Without further qualification, the quotes remain valid only for this specification. Attempts to apply the quotes to instruments with modified conventions could produce inconsistent prices. Finally, the quotes are typically provided in the form of volatility parameters of an auxiliary model used to describe the evolution of the underlying swap rate $S$. The set of quotes arranged in the form of a matrix defined by the first two dimensions is referred to as a volatility surface. Combining surfaces for different strikes leads to the volatility cube. In this sense, the cube consists of volatilities $\sigma^{T^\mathrm{E},T^\mathrm{M}-T^\mathrm{S},K}$ corresponding to a combination of an exercise date $T^\mathrm{E}$, swap tenor $T^\mathrm{M}-T^\mathrm{S}$, and strike rate $K$. Two models have become broadly established: the Black model and the Bachelier model. Due to its ability to incorporate negative swap rates, we will concentrate on the latter in this paper. 
\par Given a plain vanilla interest rate swap $\IRS(T^\mathrm{S},T^\mathrm{M}, K-S^\mathrm{p}(t), N)$ with the remaining conventions inherited from the volatility cube, the Bachelier model describes the evolution of the corresponding swap rate as 
\begin{equation*}
    \mathrm{d}S(u) = \sigma_\mathrm{B}\mathrm{d}W^C(u) \quad\text{with} \quad S(t) = S^\mathrm{p}(t)
\end{equation*}
for $u>t$, where $S^\mathrm{p}(t)$ stands for the currently observed fair swap rate and $W^C(u)$ for the Brownian motion under the annuity measure $Q^C$ associated with the annuity of the underlying swap acting as numéraire. Note that we dropped the arguments identifying the instrument from the superscript of the volatility parameter $\sigma_\mathrm{B}=\sigma^{T^\mathrm{E},T^\mathrm{M}-T^\mathrm{S},K}$ for convenience. Given these dynamics and the price formula for a swap established in \eqref{eq:irs_price_through_swap_rate}, the price of a European swaption with exercise at $T^\mathrm{E}$ and settlement date $T^\mathrm{S}$ on a swap $\IRS(T^\mathrm{S},T^\mathrm{M}, K-S^\mathrm{p}(t), N)$ is determined by
\begin{align}\label{eq:eswo_bachelier_formula}
    \ESWO^\mathrm{B}_t (T^\mathrm{E},T^\mathrm{S},\IRS) & = C(t) \EE^C \left[  \left(\omega\left( K-S(T^\mathrm{E}) \right) \right)^+ \given\iF_t \right] \nonumber \\
    & = C(t)\sigma_\mathrm{B} \sqrt{T^\mathrm{E}-t}\left[  -\omega z_t \Phi(-\omega z_t)+\phi(z_t) \right]
\end{align}
with $z_t \coloneqq  (S^\mathrm{p}(t)-K)/(\sigma_\mathrm{B}\sqrt{T^\mathrm{E}-t})$. The market prices of potential calibration instruments are then calculated by substituting the corresponding value for $\sigma_\mathrm{B}$ from the volatility cube. For completeness, we provide the derivation of \eqref{eq:eswo_bachelier_formula} in Appendix~\ref{app:bachelier_model}.
\par Henceforth, we describe the construction of various calibration sets for a Bermudan swaption $\BSWO ( \iT^\mathrm{E}, \iT^\mathrm{S}, \IRS)$ on an interest rate swap $\IRS ( \iT, \tilde{\iT}, S, L^\Delta,\iN, \tilde{\iN}, \tau, \tilde{\tau}, \omega)$ as defined in Section~\ref{sec:bswo}, valued at time $t$.
We recall that $\iT^\mathrm{E} = \{ T^\mathrm{E}_1, \cdots, T^\mathrm{E}_p\}$ and $\iT^\mathrm{S} = \{ T^\mathrm{S}_1, \cdots, T^\mathrm{S}_p\}$ denote the exercise and settlement dates, respectively, while $\iT=\{T_j\}_{j=0}^n$ and $\tilde{\iT}=\{\tilde{T}_i\}_{i=0}^m$ as well as $\iN = \{ N_1, N_2, \ldots,N_n  \}$ and $\tilde{\iN} = \{ \tilde{N}_1, \tilde{N}_2, \ldots,  \tilde{N}_m  \}$ denote the payment and notional schedules of the underlying swap, respectively. 
\par \textbf{Global Column} \newline This calibration set is fully independent of the valued product and comprises instruments with increasing times to settlement and identical tenors of the underlying swaps. The name is motivated by the fact that instruments with a constant tenor slice vertically through the volatility surface. Given a fixed tenor $T^\mathrm{TEN}$, a maximum time to settlement $T^\mathrm{SET}$, and the desired cardinality $N^\iC$, we specify the instruments forming the calibration set by 
\begin{align*}
    I_i & = \ESWO(\hat{T}^\mathrm{E}_i, \hat{T}^\mathrm{S}_i,\IRS^i) \\
    \text{with} \quad \IRS^i & = \IRS( \hat{T}^\mathrm{S}_i, \hat{T}^\mathrm{S}_i+T^\mathrm{TEN}, 0, 10^7), \\
    \hat{T}^\mathrm{S}_i & = t+i\cdot T^\mathrm{SET}/N^\iC \quad \text{for}\quad i = 1,\ldots, N^\iC.
\end{align*}
Any parameter not explicitly mentioned is inherited directly from the volatility cube. This includes the notice periods necessary to derive the exercise dates $\hat{T}_i^\mathrm{E}$. This setting ensures that the corresponding target prices $\mathrm{PV}^\mathrm{M}_i$ can be calculated using the Bachelier formula with the volatility parameters extracted from the cube. Since all the instruments have distinct exercise dates and underlying maturity dates, this approach is, in principle, suitable for calibrating both the diffusion coefficient and the mean reversion speed of the HW1F model. 
\par The choice of $T^\mathrm{TEN}$, $T^\mathrm{SET}$, and $N^\iC$ constitutes a modeling assumption. Despite the global character of the method, the choice of the first two parameter values should reflect the instruments that will be priced using the calibrated model. Since the subsequent analysis is not restricted to specific tenors or exercise dates, we simply set both parameters to half of the respective maximum values provided by the volatility cube, i.e. $T^\mathrm{SET}= 10$ and $T^\mathrm{TEN} = 15$. In the absence of other information, this choice is intended to target the central region of the ATM volatility surface. The cardinality of the calibration set will be specified later. 
\par \textbf{Global Co-terminal} \newline This calibration set is fully independent of the valued product and comprises instruments for which the time to settlement and the tenor of the underlying swap sum to a constant duration $T^\mathrm{MAT}$. Instruments constructed in this manner are referred to as co-terminal, since all underlying swaps have the same maturity date. It is worth noting that the set of core European swaptions derived from any given Bermudan swaption is also co-terminal, which provides a motivation for considering this calibration set. The approach is also referred to as diagonal calibration in the literature (see, e.g., \cite[p.~27]{haganMETHODOLOGYCALLABLESWAPS2004}), as co-terminal instruments slice diagonally through the volatility surface.
\par Given a common time to maturity $T^\mathrm{MAT}$ for the underlying swaps and the desired cardinality $N^\iC$, we specify the instruments forming the calibration set by
\begin{align*}
    I_i & = \ESWO( \hat{T}_i^\mathrm{E}, \hat{T}^\mathrm{S}_i, \IRS^i) \\
    \text{with} \quad \IRS^i & = \IRS( \hat{T}^\mathrm{S}_i, t+T^\mathrm{MAT},0, 10^7), \\
    \hat{T}^\mathrm{S}_i & = t + i\cdot T^\mathrm{MAT} / (N^\iC+1)  \quad \text{for}\quad i = 1,\ldots, N^\iC.
\end{align*}
As in the previous approach, any parameter not explicitly mentioned is inherited directly from the volatility cube. This allows us to calculate the target prices $\mathrm{PV}^\mathrm{M}_i$ using the Bachelier formula. Since all the instruments have distinct exercise dates but identical underlying maturity dates, this approach is, in principle, suitable for calibrating only the diffusion coefficient of the HW1F model. Again, the choice of $T^\mathrm{MAT}$ and $N^\iC$ is a modeling assumption. Following the same reasoning as above, we set $T^\mathrm{MAT}=T^\mathrm{SET}+T^\mathrm{TEN}=25$. The cardinality of the calibration set will be specified later. 
\par \textbf{Local Bullet ATM} \newline The calibration set depends on the valued product through the settlement dates of the core swaptions $T_1^\mathrm{S}, \cdots, T_p^\mathrm{S}$ and the maturity of the underlying swaps $T^\mathrm{M}$. The name indicates two properties. First, the approach ignores the notional schedules of the underlying swaps and employs only bullet instruments. Second, the volatility data requirements are restricted to the ATM surface. The calibration instruments are defined as follows   
\begin{align*}
    I_i & = \ESWO(\hat{T}^\mathrm{E}_i, T^\mathrm{S}_i, \IRS^i) \\
    \text{with} \quad \IRS^i & = \IRS( T^\mathrm{S}_i, T^\mathrm{M}, 0, 10^7 ) \quad \text{for}\quad i = 1,\ldots,p.
\end{align*}
Again, all parameters not explicitly mentioned are inherited directly from the volatility cube. Consequently, the corresponding target prices $\mathrm{PV}^\mathrm{M}_i$ can be calculated using the Bachelier formula. Since all the instruments have the same underlying swap maturity $T^\mathrm{M}$, this approach is, in principle, suitable for calibrating only the diffusion coefficient of the HW1F model. 
\par \textbf{Local Bullet Core} \newline This calibration set can be viewed as a refinement of the previous approach. In addition to the settlement dates of the core swaptions $T_1^\mathrm{S}, \cdots, T_p^\mathrm{S}$ and the maturity of the underlying swaps $T^\mathrm{M}$, the calibration set depends on the moneyness of the core swaptions $M_i(t)\coloneqq S-S_i^\mathrm{p}(t)$. This fact is reflected in the term \enquote{core}, which also indicates that access to the volatility cube is necessary to construct the set. The calibration instruments are defined as follows  
\begin{align*}
    I_i & = \ESWO(\hat{T}^\mathrm{E}_i, T^\mathrm{S}_i, \IRS^i) \\
    \text{with} \quad \IRS^i & = \IRS( T^\mathrm{S}_i, T^\mathrm{M}, M_i(t), 10^7 ) \quad \text{for}\quad i = 1,\ldots,p.
\end{align*}
All the comments from the previous method, in particular the calculation of the target prices $\mathrm{PV}^\mathrm{M}_i$, apply here as well. 
\par \textbf{Representative Swaption} \newline This is the first calibration set considered in this paper that accounts for the notional schedules of the underlying swap. The chosen instruments depend on the present value, delta and gamma of the underlying swaps corresponding to the core European swaptions of the Bermudan swaption. Since the notional schedules affect all three quantities, the calibration set implicitly reflects the notional evolution. We will use this calibration set only as a benchmark for the calibration sets defined subsequently. For this reason, we confine the discussion to the main idea.
\par For each core swaption, we aim to find a representative plain vanilla swaption, i.e. an instrument whose underlying matches the present value, delta, and gamma as closely as possible. Note that the sensitivities are considered with respect to the discounting curve. To achieve this, we consider the swap tenor, the constant notional, and the swap rate of the plain vanilla instrument as free parameters. At the same time, the settlement date is inherited from the core swaption. All parameters not explicitly mentioned are inherited directly from the volatility cube. Overall, this results in a system of three equations in three unknowns. To derive the calibration set, two aspects have to be considered: first, how to calculate the delta and gamma of a given interest rate swap, and second, how to solve the resulting system of equations numerically. Given the solutions $T_i^*, N_i^*$ and $S_i^*$, the calibration instruments are defined as follows 
\begin{align*}
    I_i & = \ESWO(\hat{T}^\mathrm{E}_i, T^\mathrm{S}_i, \IRS^i) \\
    \text{with} \quad \IRS^i & = \IRS( T^\mathrm{S}_i, T_i^\mathrm{S}+T_i^*, S_i^*-S^{*,\mathrm{p}}_i(t), N_i^* ) \quad \text{for}\quad i = 1,\ldots,p.
\end{align*}
Note that $S^{*,\mathrm{p}}_i(t)$ stands for the par rate of the $i$-th representative swap with tenor $T_i^*$. Again, the corresponding target prices $\mathrm{PV}^\mathrm{M}_i$ can be calculated using the Bachelier formula. The implementation in this paper follows the methodology described in \cite[pp.~42--45]{haganMETHODOLOGYCALLABLESWAPS2004}.
\par \textbf{Basket Hagan} \newline We continue with another notional-sensitive calibration set. The instruments are specified as the core European swaptions of the Bermudan swaption being priced, with the sole exception of potentially different exercise dates, i.e. 
\begin{align*}
    I_i & = \ESWO(\hat{T}^\mathrm{E}_i, T^\mathrm{S}_i, \IRS^i) \\
    \text{with} \quad \IRS^i & \coloneqq \IRS( \iT\vert_{T^\mathrm{S}_i}, \tilde{\iT}\vert_{T^\mathrm{S}_i}, S, L^\Delta,\iN\vert_{T^\mathrm{S}_i}, \tilde{\iN}\vert_{T^\mathrm{S}_i}, \tau, \tilde{\tau}, \omega) \quad\text{for}\quad i= 1, \ldots, p.
\end{align*}
The exercise dates $\hat{T}^\mathrm{E}_i$ are derived by applying the notice period and calendar convention of the volatility cube. The actual novelty of this approach stems from the derivation of the corresponding target prices $\mathrm{PV}^\mathrm{M}_i$. In what follows, we use the example of the first core swaption to adapt the procedure described in \cite[pp.~47--49]{haganMETHODOLOGYCALLABLESWAPS2004} to our setting.
\par To start with, we point out that the name of this family of approaches reflects the use of a basket of plain vanilla swaptions for price replication. The number of instruments in the basket matches the number of payments $n$ on the fixed leg of the underlying amortizing swap. All basket swaps start at the same date $T^\mathrm{S}_1$ as the amortizing swap and end at one of the fixed leg payment dates $T_1, \cdots, T_n$, respectively. The swap rates are set such that all basket swaps have the same moneyness $\lambda$, i.e. $S_j= S^\mathrm{p}_j(t)+ \lambda$ for $j=1,\ldots,n$. All parameters not mentioned explicitly are inherited from the volatility cube. Given this setting, the basic idea of the approach is to replicate the cash flows of the amortizing swap with the plain vanilla instruments from the basket. Before proceeding with the detailed derivation, it is useful to make a few remarks.   
\par Note that inheriting the conventions from the volatility cube is not arbitrary. This ensures that the prices of the basket instruments can be calculated using the Bachelier formula. At the same time, we have to account for potential differences in the conventions that could change the resulting schedules and interfere with the cash flow structure. We distinguish two relevant sources: the payment frequencies and the business day conventions. First, distinct payment frequencies may result in non-coinciding sets of payment dates. This could introduce new cash flows into the replication procedure and therefore make it less tractable. In practice, the problem is more structured because the payment intervals are usually integer multiples of one another, e.g., in the euro-denominated swaption market the fixed payment interval is almost always one of 3, 6 or 12 months. Moreover, the longest payment interval of 12 months is also the standard convention of the volatility cube. Even though this does not solve the distinct schedule issue, it ensures that the schedule of the plain vanilla swap is a subset of the schedule of the amortizing swap (subject to the business day adjustment considered subsequently). Therefore, we do not inadvertently introduce new dates into the replication problem. The situation is different for non-matching business day conventions. In practice, since mismatches are fairly rare and the resulting discrepancies do not exceed a few days, we ignore this source of inaccuracy. To ease the notation, the following calculations are shown assuming that all conventions are identical. In this sense, the simplified setting does not restrict the general applicability of the approach.   
\par The focus can now shift to the replication procedure. Recalling the price formula of a swap from \eqref{eq:pv_swap_as_bond}, the present value $\Pi$ of the defined basket of plain vanilla swaps with notionals $\omega_1, \cdots, \omega_n$ is given by  
\begin{align}\label{eq:pv_basket_plain_vanilla}
    \Pi & = \sum_{j=1}^n \omega_j \left[  P_\mathrm{c}(t,T_j)-P_\mathrm{c}(t,T_0)+\sum_{i=1}^j\tau_i\left(S^\mathrm{p}_j (t)+S^\mathrm{BS}_j+\lambda\right)P_\mathrm{c}(t,T_i) \right] \nonumber\\
    & = -\sum_{j=1}^n\omega_jP_\mathrm{c}(t,T_0)+\sum_{i=1}^n \left[ \tau_i\left( \sum_{j=i}^n \omega_j\left(  S^\mathrm{p}_j (t)+S^\mathrm{BS}_j+\lambda\right)\right)+\omega_i \right]P_\mathrm{c}(t,T_i),
\end{align}
where the order of summation was changed to obtain the representation. Comparing coefficients between \eqref{eq:pv_swap_as_bond} and \eqref{eq:pv_basket_plain_vanilla} results in the following system of $n+1$ equations
\begin{align}\label{eq:hagan_basket_lse}
    c_0 & = -\sum_{j=1}^n \omega_j \nonumber, \\
    c_j & = \omega_j + \tau_j\left( \sum_{i=j}^n \omega_i\left(  S^\mathrm{p}_i (t)+S^\mathrm{BS}_i+\lambda\right)\right)\quad \text{for}\quad j= 1, \cdots,n.
\end{align}
For fixed $\lambda$, the problem becomes a linear system of $n$ equations $\mathbf{A} (\omega_1, \cdots , \omega_n)^\mathsf{T}=(c_1, \cdots, c_n)^\mathsf{T}$, where $\mathbf{A}$ is an upper triangular matrix. Note that $\mathbf{A}$ controls the aggregation and allocation of cash flows occurring on the same dates as the corresponding coupons $c_1, \ldots, c_n$ of the amortizing underlying swap. For convenience, we define $c_{j,i}\coloneqq \tau_j ( S_i^\mathrm{p}(t)+S_i^\mathrm{BS}+\lambda)$ as the $j$-th coupon of the $i$-th basket instrument per unit notional. The matrix is then given by
\begingroup
\renewcommand{\arraystretch}{1.8}
\setlength{\arraycolsep}{5pt}
\begin{equation}
    \mathbf{A} = \begin{pmatrix}
    1+c_{1,1} & c_{1,2}   & c_{1,3} & \cdots & c_{1,n-1}     & c_{1,n} \\
    0         & 1+c_{2,2} & c_{2,3} & \cdots & c_{2,n-1}     & c_{2,n} \\
    \vdots    & \ddots    & \ddots  & \ddots &               & \vdots \\
    \vdots    &           & \ddots  & \ddots & \ddots        & \vdots \\
    \vdots    &           &         & 0      & 1+c_{n-1,n-1} & c_{n-1,n} \\
    0         & \cdots    & \cdots  & \cdots & 0             & 1+c_{n,n}
\end{pmatrix}.
\end{equation}
\endgroup
Overall, the problem can be solved efficiently using a standard root-solver to find a $\lambda$ that satisfies the first equation of \eqref{eq:hagan_basket_lse}. Practical experience indicates that the optimal $\lambda^*$ lies relatively close to the moneyness of the amortizing swap. Choosing it as the initial value of the numerical routine can accelerate convergence. 
\par Given $\omega_1, \cdots, \omega_n$, we know which plain vanilla swaps exactly replicate the cash flows of the amortizing swap. We can now use those co-initial swaps to construct a calibration set for an intermediate Hull--White model with a constant diffusion coefficient and a time-dependent mean reversion speed 
\begin{align*}
    \hat{I}_i & = \ESWO(\hat{T}^\mathrm{E}_1, T^\mathrm{S}_1, \hat{\IRS}^i) \\
    \text{with} \quad \widehat{\IRS}^i & = \IRS( T^\mathrm{S}_1, T_i, \lambda, \omega_i) \quad \text{for}\quad i = 1,\ldots,n.
\end{align*}
The corresponding target prices $\smash{\widehat{\mathrm{PV}}}_i^\mathrm{M}$ can be calculated using the Bachelier formula. A calibration set in which all the instruments have the same settlement date is often referred to as row calibration in the literature (see, e.g., \cite[pp.~22--23]{haganMETHODOLOGYCALLABLESWAPS2004}). Given these prices, we can calibrate the mean reversion speed of the intermediate model for an arbitrary but realistic $\sigma$ (e.g., aligned with the quoted market volatility of the plain vanilla swaptions). The resulting HW1F model can then be used to obtain the price of the amortizing European swaption $\mathrm{PV}_1^\mathrm{M}$. Repeating the above steps for all core swaptions completes the main calibration set.  
\par \textbf{Basket Naive} \newline This method is a modification of the previously presented \enquote{Basket Hagan} approach. We first encountered the underlying approximation in a practical implementation. It was also outlined in practitioner presentations held at QuantLib User Meetings by Miemiec~\cite{miemiec2013accretingSwaptions} and Kürzinger~\cite{kuerzinger2017aspectsIrregularSwaptions}. An implementation of the underlying approximation can also be found in the experimental module of the QuantLib finance library~\cite[see,~e.g.,][version~1.44]{quantlib}. However, we found no systematic assessment of the suitability of the approximation for Bermudan swaption pricing, which motivated us to investigate its properties and the consistency of the resulting valuations. The approach addresses a practical disadvantage of existing basket approaches presented in the literature, namely the need for an intermediate mean-reversion calibration. In a typical pricing procedure for a Bermudan swaption, a substantial portion of the runtime is spent on the numerical calibration of model parameters. Introducing a repeated need for a mean-reversion calibration further increases the computational cost of an already expensive step.
\par To eliminate the need for intermediate model instances, we first consider the present value of a general amortizing swap expressed as a sum of plain vanilla swaps using the Hagan decomposition introduced above
\begin{equation*}
    \IRS_t^\mathrm{AMO} = \sum_{i=1}^n \widehat{\IRS}^i_t.
\end{equation*}
Taking conditional expectation of the discounted payoff to derive the price of the corresponding amortizing European swaption yields
\begin{equation}
    \ESWO_t^\mathrm{AMO} = \EE^Q \left[ \exp\left(-\int_t^{T^\mathrm{E}}r(u)\dd u\right)  \left(\sum_{i=1}^n \widehat{\IRS}_{T^\mathrm{E}}^i\right)^+ \given \iF_t \right] \le\sum_{i=1}^n \widehat{\mathrm{PV}}^i_t.\label{eq:upper_bound_approximation}
\end{equation}
Due to the subadditivity of the positive-part function, changing the order of summation and the positive-part operator results in an inequality. However, if the upper-bound approximation is sufficiently tight, it can be used as the target price $\mathrm{PV}^\mathrm{M}$ for the calibration of the main model, effectively eliminating the need for the intermediate mean-reversion calibration. Whether the upper-bound gap is sufficiently small to be used for the calibration of the diffusion coefficient is a central question investigated in the empirical study presented in the next section.
\par A few remarks on why we expect this approximation to be tight enough are in order. We point out that the following arguments are purely qualitative and do not constitute a formal justification. First, from the perspective of the valuation date, all the basket instruments have an identical moneyness level $\lambda$. If the exercise decision were made immediately, all basket underlyings would have the same sign. Clearly, this does not guarantee that the options will be simultaneously in or out-of-the-money at the joint exercise date. However, for larger absolute values of $\lambda$ and sufficiently correlated swap rates, this seems to be more likely. Second, the basket notionals $\omega_1, \ldots, \omega_n$ act as weights of individual instruments. Their distribution depends on many factors and is often highly unequal. This suggests that in some cases the value of the basket could be materially driven by only a few instruments or, in extreme cases, a single instrument.
\par Third, Hunt and Kennedy~\cite{huntImpliedInterestRate1998} showed that, given a calibrated one-factor short-rate model, one can choose the basket instruments in such a way that the equality in \eqref{eq:upper_bound_approximation} holds. The general specification of the basket instruments is identical to the one in the Hagan method presented above. However, their approach applies the Jamshidian decomposition to find the critical rate of the amortizing swap and sets the strikes of the basket instruments to yield identical critical rates. This ensures that, at the exercise date, all contributing underlying swap values have the same sign in all possible states. Alternatively, to examine how tight the upper-bound approximation is, we could focus on the differences between critical rates implied by the strikes in the Hagan approach. This relationship between the upper-bound gap and the dispersion of the critical rates will be investigated in the empirical part of the paper. 
\par At this point, we present a slightly different perspective and compare corresponding instruments across both baskets directly. To this end, we assume that the Hunt-Kennedy strikes $\tilde{S}_1,\ldots,\tilde{S}_n$ and weights $\tilde{\omega}_1, \ldots, \tilde{\omega}_n$ are known and start the investigation with the longest instruments. If their rates match, the resulting notionals have to be identical because both instruments generate the same last cash flow $c_n$ from \eqref{eq:hagan_basket_lse}. To illustrate the other case, we assume that the Hagan rate is significantly higher, i.e. $S_n = \tilde{S}_n+d$ with $d>0$. To match the coupon $c_n$, the higher rate has to be compensated by a smaller notional $\omega_n$. This relationship is governed by the following equation
\begin{equation}
\frac{\omega_n}{\tilde{\omega}_n} = \frac{1+\tau_n\tilde{S}_n}{1+\tau_n(\tilde{S}_n+d)}.\label{eq:hagan_hunt_kennedy_matching}
\end{equation}
For any reasonable choice of $\tilde{S}_n$ and $d>0$, the right-hand side is close to 1, resulting in only a moderate notional reduction. In this example, the potential price difference between the longest basket instruments will be mainly driven by the change in the strike. Without further assumptions regarding the deal characteristics, it is not possible to infer how this will influence the present value. However, since both baskets replicate all cash flows simultaneously, the change in the last period has to be compensated by an opposing movement in rates and notionals across some of the shorter instruments. Recall that reducing one weight has to increase the sum of the remaining ones. Otherwise, the first equation of \eqref{eq:hagan_basket_lse} is violated. Moreover, despite the lower notional, the longest instrument contributes higher interest payments. Since all the cash flows have to be matched simultaneously, this has to be compensated by lower strikes across some of the shorter instruments. We do not claim that those opposite movements will result in price differences of exactly the same magnitude. We only emphasize that price differences between corresponding basket instruments are expected to have different signs, possibly introducing a material mitigating effect.   
\par Finally, Roos~\cite{roos2016ModelIndependentBounds} investigated a similar approximation for amortizing Bermudan swaptions decomposed into a basket of bullet Bermudan swaptions. While the results suggest that the bounds are tight enough to provide meaningful tests for arbitrage or even a potential static hedge strategy, the bounds were not explicitly considered as an alternative pricing approach. Recall that the price of a Bermudan swaption is often decomposed into two components: the value of the first European exercise right and the so-called switch value representing the additional value of the right to postpone the exercise decision into the future. From this perspective, the approximation problem for a European swaption could turn out to be less complex since it avoids additional effects resulting from different optimal exercise strategies of individual basket options. Before turning to the calibration algorithm, we conclude this subsection with Table~\ref{tab:comparison_calibration_methods} summarizing the essential characteristics of all methods presented previously.
\begin{table}[t]
\small
\centering
\begin{threeparttable}
\caption{\textbf{Comparison of calibration-set construction methods.}}
\label{tab:comparison_calibration_methods}
\begin{tabular*}{\textwidth}{@{\extracolsep{\fill}}lccccccc}
\toprule
Aspect 
& \makecell{Global\\Co-terminal}
& \makecell{Global\\Column}
& \makecell{Local\\ATM}
& \makecell{Local\\Core}
& \makecell{Rep.\\Swaption}
& \makecell{Basket\\Hagan}
& \makecell{Basket\\Naive} \\
\midrule
Applicable to $a(t)$-calibration             & N & Y & N & N & $\text{N}^1$ & N & N \\
Applicable to $\sigma (t)$-calibration        & Y & Y & Y & Y & Y            & Y & Y \\
Notional sensitivity                      & N & N & N & N & Y            & Y & Y \\
Market data requirements                  & ATM & ATM & ATM & cube & cube & cube & cube \\
$\text{Computational cost}^2$             & \sminus & \sminus & \lminus & \neutral & \lplus & \splus & \lplus \\
$\text{Implementation complexity}^2$      & \lminus & \lminus & \neutral & \neutral & \lplus & \splus & \lplus \\
Non-standard conventions                  & N  & N & N & N & Y & Y & Y \\
\bottomrule
\end{tabular*}
\begin{tablenotes}[online,flushleft]
\footnotesize
\item[]\hspace*{-\labelsep}\textit{Notes:} 1) Not in general because the derived representative swaptions could theoretically have the same maturity. 2) Relative score with \enquote{Local Bullet Core} as baseline: \sminus, \lminus, \neutral, \lplus, \splus. 
\end{tablenotes}
\end{threeparttable}
\end{table}
\subsubsection{Calibration algorithm}
\par The main goal of the calibration algorithm is to find the values of the free parameters such that the market prices are replicated by the model as closely as possible. As discussed in Section~\ref{sec:hw1f_model}, we employ two HW1F model specifications: piecewise constant mean reversion speed $a(t)$ with exogenously specified constant diffusion coefficient $\sigma(t)\equiv\sigma$ and, conversely, piecewise constant diffusion coefficient $\sigma(t)$ with exogenously specified constant mean reversion speed $a(t)\equiv a$. Before formally defining the problem, a few comments on how prices of European swaptions depend on both functions are in order. 
\par A key observation is that, in the HW1F model, the price of a European swaption with exercise date $T^\mathrm{E}$ and underlying swap maturity at $T^\mathrm{M}$ depends on the diffusion coefficient $\sigma(t)$ over the interval $[t,T^\mathrm{E}]$ and the mean reversion speed $a(t)$ over the interval $[t,T^\mathrm{M}]$. Since the price of a European swaption can be decomposed into a sum of zero-coupon bond option prices, this dependence can be seen directly from the price formula for $\ZBO$ stated in \eqref{eq:zbo_pricing_formula}. The diffusion coefficient enters only through the integral $V(\cdot,\cdot)$ defined in \eqref{eq:hw1f_drift_term_result} with the upper limit $T^\mathrm{E}$. The mean reversion speed also enters through the bond-loading function $B(\cdot,\cdot)$ defined in \eqref{eq:hw1f_zero_coupon_bond_formula} by an integral with upper limit $T^\mathrm{M}$. It is also worth noting that the price of the zero-coupon bond option, and therefore the European swaption, is continuous in the step values of the mean reversion speed as well as the diffusion coefficient. The piecewise constant construction of both functions does not affect the continuity, since both enter the pricing formula through integrals. Moreover, European swaption prices are non-decreasing in a single step value $\sigma_i$ of the diffusion coefficient $\sigma(t)$ while all other steps are fixed.
\par With this in mind, we can sort the calibration instruments by exercise dates $T^\mathrm{E}_1,\ldots,T^\mathrm{E}_n$ or underlying maturities $T^\mathrm{M}_1,\ldots,T^\mathrm{M}_n$ for $\sigma (t)$ and $a(t)$, respectively. This results in a structure in which the model price of the $i$-th instrument depends only on the first $i$ steps of the calibrated parameter $\theta\in\{a,\sigma\}$, i.e. $\mathrm{PV}_i^\mathrm{HW1F}=\mathrm{PV}_i^\mathrm{HW1F}(\theta_1,\ldots,\theta_i)$. This allows us to define a sequence of one-dimensional problems in which each calibration instrument determines one additional parameter value by setting the pricing residual $f_i\coloneqq \mathrm{PV}^\mathrm{HW1F}_i(\theta_i;\theta_1, \cdots,\theta_{i-1} ) - \mathrm{PV}^\mathrm{M}_i$ to zero. More precisely, we are looking for a vector $\theta =(\sigma_1,\ldots,\sigma_n)$ or $\theta = (a_1,\ldots, a_n)$ corresponding to the intervals induced by $t<T^\mathrm{E}_1<\ldots<T^\mathrm{E}_n$ or $t<T^\mathrm{M}_1<\ldots<T^\mathrm{M}_n$, respectively. Formally, we aim to find 
\begin{equation*}
    \theta_i \in [L^\theta, U^\theta] \quad \text{such that} \quad f_i= 0 \quad \text{for}\quad i= 1, \ldots,n. 
\end{equation*}
\par The choice of the bounds $[L^\theta, U^\theta]$ will be discussed later. The numerical method needed to solve the derived problems is outside the scope of this paper. We employ a standard numerical routine provided by our computing environment. The chosen approach depends on the calibrated parameter $\theta$. Due to the monotonicity and continuity of the model prices $\mathrm{PV}^\mathrm{HW1F}$ with respect to a single step value of the diffusion coefficient, a root-finding method is employed. The calibration of the mean reversion speed is treated as an optimization problem. For the remainder of this subsection, we consider the general procedure specified in Algorithm~\ref{alg:calibration_sigma_hw1f}. In particular, we focus on diagnostics that may provide insights into the quality of the resulting parameters and whether the instruments provide additional calibration information.   
\begin{algorithm}[!t]
\caption{Sequential calibration of a stepwise constant HW1F parameter}
\label{alg:calibration_sigma_hw1f}
\begin{algorithmic}[1]
\Statex \textbf{Input: } calibration set $\iC$, parameter $\theta$, bounds $[L^\theta,U^\theta]$
\Statex \textbf{Output: } diagnostics, parameter values $\theta = (\theta_1, \cdots, \theta_n)^\textsf{T} $ or failure
\State Extract jump dates of the free parameter $\theta$ from the instruments
\State Sort the instruments in ascending order with respect to the extracted jump dates
\For{$i = 1$ to $n$}
    \Statex \hspace{\algorithmicindent}\textbf{--- Calibration}
    \If{$\theta = \sigma$}
        \State Evaluate the root-existence flag, i.e. $f_i(L^\sigma) \cdot f_i(U^\sigma) \le 0$  \State Evaluate the low-sensitivity flag, i.e. $\vert f_i(L^\sigma) -f_i(U^\sigma)\vert / \mathrm{Notional}_i < \qty{3}{\bp}$
        \If{root exists}
            \State Find $\theta_i$ solving $f_i(\theta_i)=0$
        \EndIf
    \ElsIf{$\theta = a$}
        \State Find $\theta_i$ minimizing $f_i^2(\theta_i)$
    \EndIf
    \Statex \textbf{--- Fallback handling}
    \If{$\theta =\sigma$}
        \If{solution $\theta_i$ found}
            \State Evaluate bound-hit flag, i.e. $\vert L^\theta - \theta_i\vert<\qty{1}{\bp}$ or $\vert U^\theta - \theta_i\vert<\qty{1}{\bp}$
        \EndIf
        \State $C \gets
            \text{no root exists} \lor \text{numerical routine fails} \lor \text{(bound-hit} \land \text{low-sensitivity)}$
    \ElsIf{$\theta =a$}   
            \State $C \gets \text{numerical routine fails}$
    \EndIf
    \If{$C$}
        \If{$i=1$}
            \State Merge the first two intervals
            \State Set initial-merge flag
            \State \textbf{continue}
        \ElsIf{$i=2$ and initial-merge flag is set}
            \State Terminate and return error
        \Else
            \State Use the previous interval's value as fallback, i.e. $\theta_i = \theta_{i-1}$
        \EndIf
    \EndIf
    \If{$i=2$ and initial-merge flag is set}
        \State Set $\theta_1=\theta_2$
        \State Evaluate post-calibration diagnostics for the first instrument
    \EndIf
    \Statex \textbf{--- Post-calibration diagnostics}
    \State Reevaluate the bound-hit flag, i.e. $\vert L^\theta - \theta_i\vert<\qty{1}{\bp}$ or $\vert U^\theta - \theta_i\vert<\qty{1}{\bp}$
    \State Evaluate the fit-warning flag, i.e. $\vert f_i(\theta_i)\vert /\mathrm{Notional}_i> \qty{0.00001}{\bp}$
\EndFor
\end{algorithmic}
\end{algorithm}
\par To begin with, the rationale behind the choice of the bounds differs depending on the calibrated parameter. For the diffusion coefficient $\sigma$, we set $L^\sigma= \qty{0}{\bp}$ and $U^\sigma= \qty{400}{\bp}$, heuristically motivated by the minimum and maximum quotes observed in the volatility cube. This aims at ensuring market-realistic values consistent with a broad range of model prices. Turning to $a(t)$, the primary goal of the intermediate model with time-dependent mean reversion speed is price replication. The resulting short-rate evolution has only an indirect influence on the final price of the Bermudan swaption and is used to derive synthetic prices for core amortizing European swaptions. Therefore, we set the bounds at $L^a = \qty{-100}{\percent}$ and $U^a = \qty{100}{\percent}$ to enable a broad range of possible prices.
\par Next, for each instrument we record post-calibration diagnostics in the form of Boolean flags. Together with the fitted parameter value, these flags help us assess the quality of the calibration. First, we check whether the solution $\theta_i$ is close to the respective bounds (absolute distance less than \qty{1}{\bp}), which could indicate structural misspecification, e.g., due to restrictive bounds or incompatible pricing targets. In addition, we examine whether the pricing residual is sufficiently close to zero (difference normed by the initial notional of the instrument less than $10^{-5}\,\text{bp}$).\footnote{The value is set to activate the flag if the calibration residual exceeds $10^{-2}$ units of currency for an instrument with a baseline notional of $10^7$ units of currency.} 
\par The continuity and monotonicity of the model prices with respect to an individual step value $\sigma_i$ allow additional diagnostics. To begin with, by checking the signs of the residual $f_i$ at the two endpoints $L^\sigma$ and $U^\sigma$, we can verify the existence of a root within the admissible bounds. If no root exists, this could again indicate structural misspecification. Moreover, the difference $\vert f_i(L^\sigma)-f_i(U^\sigma)\vert$ gives us the maximal achievable price range of the calibration instrument within the bounds. Dividing the term by the notional of the instrument and expressing it in basis points defines a metric that can be used to assess the sensitivity of the model price with respect to the calibrated step value. If the resulting value lies below a certain threshold, it could indicate that the calibrated step value $\sigma_i$ is weakly identified by the calibration instrument. Consequently, the potential additional market information provided by that instrument could be limited. To minimize the risk of false-positive flags, the threshold is set at a level of \qty{3}{\bp}.
\par Finally, we allow the algorithm to fall back to a previously calibrated parameter value, i.e. $\theta_i = \theta_{i-1}$. In the general case, the rule is triggered when the numerical routine fails. For the diffusion coefficient, the fallback is additionally applied if the root does not exist or the boundary-hit and low-sensitivity flags are activated simultaneously. Special treatment is applied when the fallback mechanism is triggered for the initial step value $\theta_1$. In this case, we merge the first two intervals and continue with the second instrument. If the fallback is triggered again, the algorithm terminates with failure. Otherwise, we use the calibrated value for both intervals, i.e. $\theta_1=\theta_2$, and evaluate the post-calibration diagnostics for the first instrument.   
%
%
%
%
%
\section{Empirical Results}\label{sec:empirical_results}
\par All experiments presented in this paper were conducted using EUR interest-rate market data from 1 December 2021 to 28 November 2025. The dataset comprises three yield curves associated with the €STR, three-month EURIBOR and six-month EURIBOR reference rates, as well as a volatility cube covering exercises from 3 months to 20 years, underlying swap tenors from 1 to 30 years and moneyness ranging from \qtyrange{-400}{+400}{\bp} relative to the ATM rate. We retrieved all market data from Bloomberg.
\par The zero-curve bootstrapping used linear interpolation between the logarithms of discount factors with constant extrapolation of instantaneous forward rates (see, e.g., \cite[pp.~95--96]{haganInterpolationMethodsCurve2006}). To construct the volatility cube, linear interpolation in total variance was applied along the exercise dimension and linear interpolation was used along the tenor dimension. To derive missing strike points, the SABR model was calibrated to market smiles for each combination of the exercise and tenor grid points. The implied volatilities were quoted in terms of the Bachelier model for European-style physically settled contracts. We refer to \cite[pp.~15--19]{bloombergVCUB} for the general interpolation procedure and \cite{hagan2002ManagingSmileRisk} for further details on the SABR model.\footnote{For transparency, we note that in \cite[pp.~15--19]{bloombergVCUB} Bloomberg proposes straight bilinear interpolation in the exercise and tenor dimensions. In equity and forex option pricing, interpolation along the exercise dimension in total variance is a standard technique that helps avoid calendar arbitrage. We extended this approach to the swaption cube. For the necessary multi-dimensional conditions for arbitrage-free interpolation of the swaption volatility cube, we refer to \cite{johnson2009ArbitragefreeConstructionSwaption}.}  
\par Although the experiments described below utilized different datasets of randomly generated swaptions, general market conventions (e.g., day count fraction, business day adjustments) were consistently inherited from the EUR volatility cube. This holds particularly true for the payment frequencies of the underlying swaps. We adopted annual payments for the fixed leg. For the floating leg, the frequency was set depending on the swap tenor: quarterly up to and including one year and semiannual beyond that. Option strikes and exercises covered the complete range supplied by the cube. To ensure market-oriented product design, the lower strike boundary was adjusted based on the considered market data scenario such that the resulting absolute strike fulfilled the IRRBB maturity-dependent floor defined by the EBA (see \cite[Article 3(7)]{2023CommissionDelegatedRegulation}). Amortization was applied annually with rates ranging from \qtyrange{1}{5}{\percent}, consistent with typical principal repayment rates for mortgages. The swap tenors and exercise frequencies will be specified for each dataset separately. Note that each randomly chosen product characteristic was drawn independently from a uniform distribution over its corresponding range. This aimed to establish a broad yet market-consistent sample. The calibration of the piecewise constant diffusion coefficient was based on an exogenously specified mean reversion speed of \qty{3}{\percent}. All pricing results are expressed in basis points of the initial notional.
\par\subsection{Cardinality of the calibration set}
\par The goal of the first experiment is to investigate the influence of the cardinality of the calibration set on the pricing consistency of the Bermudan swaptions. To this end, we compare valuation results based on cardinalities of 5, 10, 15, and 20 instruments to the price derived with the calibration set consisting of all core European swaptions. The sets with a reduced number of instruments always comprise the shortest and longest core exercises. The remaining instruments are distributed uniformly across all available exercise dates. All depicted calculations are based on the \enquote{Local Bullet Core} method.
\par We conduct the comparison on two independent product sets representing distinct configurations in which high cardinality may arise: high exercise frequency with medium to long underlying tenors (quarterly exercise dates with tenors ranging from 5 to 30 years) and medium exercise frequency with long underlying tenors (annual exercise dates with tenors ranging from 20 to 30 years). Each set consists of \num{1000} randomly sampled Bermudan swaptions. To cover a broad spectrum of market data scenarios, we divide both product groups into 100 subsets of 10 swaptions, evaluating each subset on a randomly selected valuation date. 
\par The results indicate that increasing the cardinality beyond 15 instruments has only marginal effects on the present values of the investigated Bermudan swaptions. For the calibration set containing 15 instruments, over \qty{95}{\percent} of observed absolute deltas remain below \qty{1}{\bp} in both product groups. Furthermore, the investigation suggests that the complete calibration set sharply increases the number of calibration flags in the dataset with frequent exercise rights. At the same time, a lower cardinality translates into faster calibration, reducing the overall pricing runtime. In the most computationally demanding case of quarterly exercise dates, restricting the calibration set to 15 instruments reduces the calibration runtime by roughly \qty{92}{\percent} compared to the unrestricted case.
\begin{table}[tbp]
\centering
\setlength{\tabcolsep}{4.5pt}
\begin{threeparttable}
\caption{\textbf{Effect of calibration-set cardinality on computational performance, calibration stability and pricing consistency.}}
\label{tab:cardinality_experiment}
\begin{tabular}{llrrrrrrrrrrrr}
\toprule
\multirow{2}{*}{Frequency} & \multirow{2}{*}{\#} & \multicolumn{4}{c}{Calibration} & \multicolumn{8}{c}{Pricing} \\
\cmidrule(lr){3-6}
\cmidrule(lr){7-14}
& & $\overline{\Delta t}$& $p_{\mathrm{flag}}$ & $\overline{\vert\Delta\sigma\vert}$  & $\overline{J}^\mathrm{m}_{\sigma}$ & Min & $P_1$ & $\overline{\Delta}$ & $P_{50}$ & $P_{99}$ & Max & $p_{<\qty{1}{\bp}}$ & $\overline{\vert\Delta\vert}_{\ge\qty{1}{\bp}}$ \\
\midrule
annual & 5 & 0.89 & 0.0 & 0.2 & 21.3 & -51.1 & -17.1 & -2.4 & -1.7 & 3.5 & 8.4 & 30.5 & 3.9 \\
 & 10 & 0.73 & 0.0 & 0.2 & 20.5 & -15.0 & -4.8 & -0.6 & -0.3 & 0.9 & 1.8 & 81.2 & 2.1 \\
 & 15 & 0.54 & 0.0 & 0.1 & 25.5 & -5.5 & -2.6 & -0.2 & -0.1 & 0.4 & 1.1 & 95.6 & 2.0 \\
 & 20 & 0.30 & 0.4 & 0.0 & 32.6 & -9.0 & -1.4 & -0.1 & -0.0 & 0.3 & 0.9 & 98.2 & 2.4 \\
 & $\infty$ & - & 0.4 & - & 34.1 & - & - & - & - & - & - & - & -\\
\addlinespace
quarterly & 5 & 0.98 & 0.0 & 0.4 & 19.6 & -69.8 & -17.1 & -1.5 & -0.5 & 2.3 & 4.3 & 53.8 & 3.7 \\
 & 10 & 0.95 & 0.8 & 0.4 & 23.9 & -25.9 & -6.0 & -0.4 & -0.1 & 0.7 & 2.9 & 87.6 & 2.5 \\
 & 15 & 0.92 & 5.1 & 0.3 & 28.4 & -11.4 & -2.5 & -0.2 & -0.0 & 0.4 & 1.3 & 95.8 & 2.2 \\
 & 20 & 0.88 & 13.7 & 0.3 & 31.2 & -4.2 & -1.4 & -0.1 & -0.0 & 0.2 & 0.5 & 98.1 & 1.9 \\
 & $\infty$ & - & 82.2 & - & 55.1 & - & - & - & - & - & - & - & -\\
\bottomrule
\end{tabular}
\begin{tablenotes}[flushleft]
\footnotesize
\item[]\hspace*{-\labelsep}\textit{Notes:} $\overline{\Delta t}$ denotes the relative reduction in mean calibration runtime compared to the unrestricted case, reported as a fraction; $p_{\mathrm{flag}}$ is the share of valuation runs with at least one calibration flag (see Algorithm~\ref{alg:calibration_sigma_hw1f}), expressed as a percentage; $\overline{\vert\Delta\sigma\vert}$ is the mean absolute volatility difference compared to the unrestricted case, expressed in basis points; and $\overline{J}^\mathrm{m}_{\sigma}$ is the mean maximum absolute jump between consecutive volatility intervals, expressed in basis points. All valuation results are presented in basis points relative to the initial notional of the swaption. The unrestricted cardinality acts as a baseline for all deltas. The final two columns report the share of absolute deltas less than \qty{1}{\bp} $p_{<\qty{1}{\bp}}$, expressed as a percentage, and the mean absolute difference conditional on $|\Delta|\ge\qty{1}{\bp}$, expressed in basis points.
\end{tablenotes}
\end{threeparttable}
\end{table}
\par The detailed results are summarized in Table~\ref{tab:cardinality_experiment}, which presents metrics for each investigated cardinality grouped into two categories: calibration and pricing. Regarding the former, we report the following metrics: the relative reduction in the mean calibration runtime compared to the unrestricted case $\overline{\Delta t}$, expressed as a fraction; the share of valuation runs with at least one calibration flag $p_\mathrm{flag}$, expressed as a percentage; the mean absolute volatility difference compared to the unrestricted case $\overline{\vert\Delta\sigma\vert}$, expressed in basis points; and the mean maximum absolute jump between adjacent volatility steps $\overline{J}^\mathrm{m}_{\sigma}$, expressed in basis points. To determine the volatility difference, we first calculate the mean volatility level for each calibration run individually (time-weighted mean) and compare the resulting values across different cardinalities for each product. For the pricing category, we provide the following statistics expressed in basis points: extreme values, quantiles, the mean, and the mean absolute delta for differences greater than or equal to \qty{1}{\bp} $\overline{\vert\Delta\vert}_{\ge\qty{1}{\bp}}$. Additionally, we present the share of absolute deltas less than \qty{1}{\bp} $p_{<\qty{1}{\bp}}$, expressed as a percentage.
\par Focusing first on the calibration procedure itself, the runtimes depend linearly on the effective number of calibration instruments. This leads to increased computational costs, particularly for options with quarterly exercise dates evaluated using the complete calibration set. Taking all core swaptions into account means that the volatility function consists of up to 120 steps for one product. Compared to the mid-sized case of 15 instruments, this results in calibration runtimes higher by a factor of over 10. Although not as pronounced, the effect is also clearly observable in the product set with annual exercise dates, resulting in runtimes roughly twice as long. Moreover, higher cardinalities correspond to a more frequent occurrence of calibration flags. Except for the quarterly exercise dates with unrestricted cardinality, where the majority of the valuation runs are affected, the effect is mild with a maximum of roughly \qty{14}{\percent} of runs for all other cases. The majority of the flags are triggered by a low sensitivity of the instrument and the no-root check. Interestingly, the maximum absolute price error in each valuation run, resulting from the volatility being carried forward from the immediately preceding interval, is relatively small with a mean of \qty{0.27}{\bp} and a maximum value of \qty{1.8}{\bp} over both product sets and all cardinalities. This suggests only limited overall pricing impact after the fallback. 
\par Turning to the calibrated volatility, the observed overall volatility level remains reasonably constant for all calibration sets, with slightly higher differences for quarterly exercise frequencies. On the other hand, the mean maximum jump tends to increase with the number of instruments. This may suggest that additional instruments do not change the mean level of the calibrated diffusion coefficient, but tend to introduce more pronounced localized movements. This is consistent with the expectation that shorter periods between exercise dates may require volatility spikes to reproduce the target market prices. 
\par We next consider pricing results depicted in Table~\ref{tab:cardinality_experiment}. First, we observe that the distributions of price differences are left-skewed and highly concentrated around zero. For calibration sets with at least 10 instruments, the mean error ranges from \qtyrange{-0.6}{-0.1}{\bp} and the median from \qtyrange{-0.3}{0}{\bp}. Valuations obtained using smaller calibration sets tend to slightly underestimate the Bermudan prices on average. A second observation concerns the overall pricing consistency. For calibration sets with 15 and 20 instruments, over \qty{95}{\percent} and \qty{98}{\percent} of the absolute deltas, respectively, are less than \qty{1}{\bp}. Since the chosen threshold is very conservative in the context of amortizing Bermudans, the results indicate that a careful restriction of the cardinality can be a viable modeling choice. With respect to the tails, the remaining absolute deltas are of the same order of magnitude. With means around \qty{2}{\bp}, they remain reasonably close to the threshold. 
\begin{figure}[tbp]
    \centering
    \includegraphics[width=\textwidth]{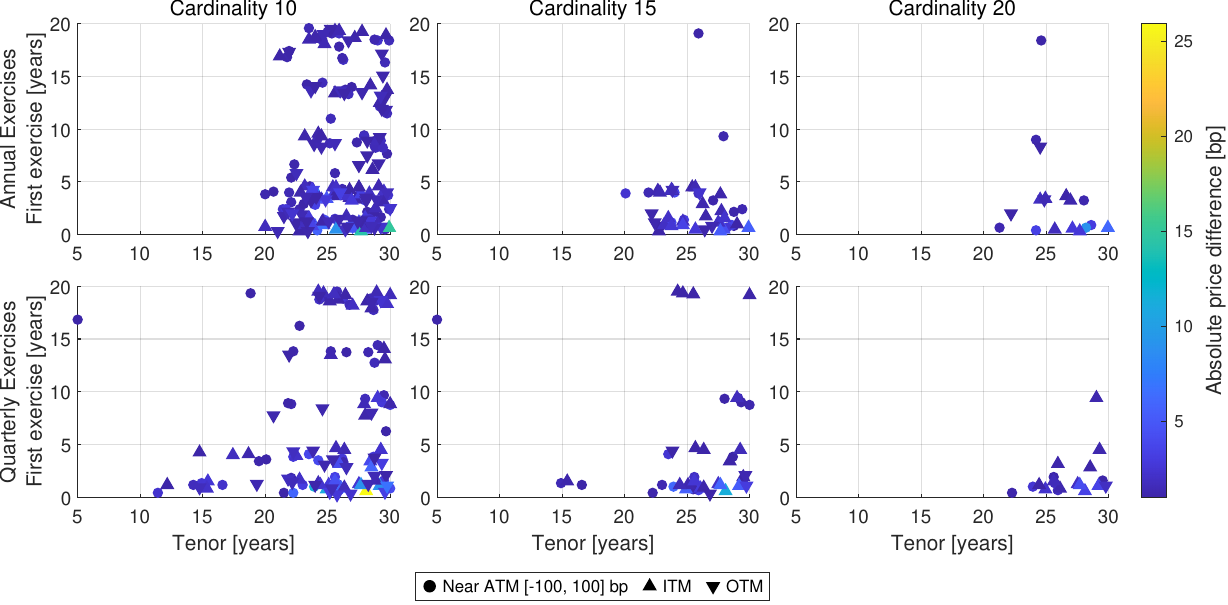}
    \caption{\textbf{Distribution of large absolute price differences -- Calibration-set cardinality.} Absolute deltas of at least \qty{1}{\bp} are mapped against the exercise of the option and the tenor of the underlying swap. Each panel shows one combination of exercise frequency and calibration-set cardinality. Point colors correspond to the magnitude of the differences. The shapes of the markers indicate the moneyness of the Bermudan swaption.}
    \label{fig:cardinality_deltas_greater_1bp}
\end{figure}
\par In addition, Figure~\ref{fig:cardinality_deltas_greater_1bp} shows scatter plots of absolute deltas greater than or equal to \qty{1}{\bp}. It specifically illustrates the relationship between the size of the delta, the first exercise date of the option and the tenor of the underlying swap. From left to right, with increasing cardinality, we observe a concentration of points in the southeast section of each panel, corresponding to shorter expiries and longer tenors. Since longer tenors imply more exercise dates, a calibration set with a fixed number of instruments covers a smaller share of the available core swaptions, which may explain the clustering in this region. Regarding the concentration at shorter expiries, we first note that the selection of instruments in incomplete calibration sets leads to equidistant spacing between exercise dates. In this respect, a given time interval between exercises is larger in relative terms at the short end of the maturity structure. Accordingly, excluding an exercise date between 10Y and 11Y seems to be less problematic than between 1Y and 2Y. This effect could be amplified if the curvature of the volatility structure is more pronounced at the short end. Additionally, both effects can reinforce one another, since longer tenors have higher annuities which can magnify small volatility mismatches. Based on the rarity and size of the absolute deltas, together with the fact that higher pricing uncertainties are generally to be expected at extreme combinations of expiries and tenors, we consider the observed differences immaterial.
\par Taken together, the results indicate that a carefully restricted cardinality of the calibration set does not materially affect pricing consistency. The marginal benefit of increasing the cardinality reaches a plateau at 15 instruments. In this case, over \qty{95}{\percent} of all resulting prices differ by less than \qty{1}{\bp}, with a mean of around \qty{2}{\bp} for absolute deltas greater than or equal to \qty{1}{\bp}. Observations beyond \qty{3}{\bp} are rare and do not exceed \qty{12}{\bp}. Furthermore, they occur only at short expiries in combination with longer tenors, a product cluster with high pricing uncertainty. This suggests that reducing the cardinality may provide a reasonable compromise between consistency, stability and computational efficiency. Based on these findings, we restrict all calibration sets, regardless of the applied construction method, to 15 instruments for the remainder of this paper.
\subsection{Naive basket replication of amortizing European swaptions}
\par The objective of the following experiment is twofold. First, we investigate whether the naive approach to basket replication can produce a tight upper bound for the prices of amortizing European swaptions. Second, we seek to determine whether the bound is tight enough to be used for pricing amortizing Bermudan swaptions. For this purpose, we initially compare present values of European swaptions calculated by naive summation of prices of the individual basket instruments with those computed using the calibrated intermediate instance of the HW1F model. Subsequently, we benchmark prices of Bermudan swaptions evaluated with the Hull--White model calibrated to Hagan prices against the results derived from the model calibrated to the naive prices. If pricing differences prove to be immaterial, the approach could eliminate the computationally expensive intermediate mean reversion calibration runs.
\par We perform the first comparison on a set of \num{10000} randomly generated amortizing European swaptions with tenors ranging from 15 months to 30 years. Similar to the previous analysis, the set is divided into 250 subsets of 40 products, each randomly assigned a valuation date to cover a broad range of market data environments. For the Bermudan benchmark study, the test set comprises \num{1000} amortizing Bermudan swaptions sampled over the same tenor range. Either a quarterly or an annual exercise frequency is chosen at random. Again, we divide the products into 100 subsets, each randomly assigned to a valuation date.        
\par The findings suggest that the upper bounds obtained by naive basket aggregation are sufficiently tight to be directly used in the calibration of the Hull--White volatility function. In over \qty{94}{\percent} of the investigated cases, the price differences for European swaptions are less than \qty{1.5}{\bp}. The share remains stable for Bermudan swaptions, providing further evidence of a low impact on pricing. In both tests, extreme deviations do not exceed \qty{5}{\bp} and are primarily observed for long tenors.         
\par
\begin{figure}[tbp]
    \centering
    \includegraphics[width=\textwidth]{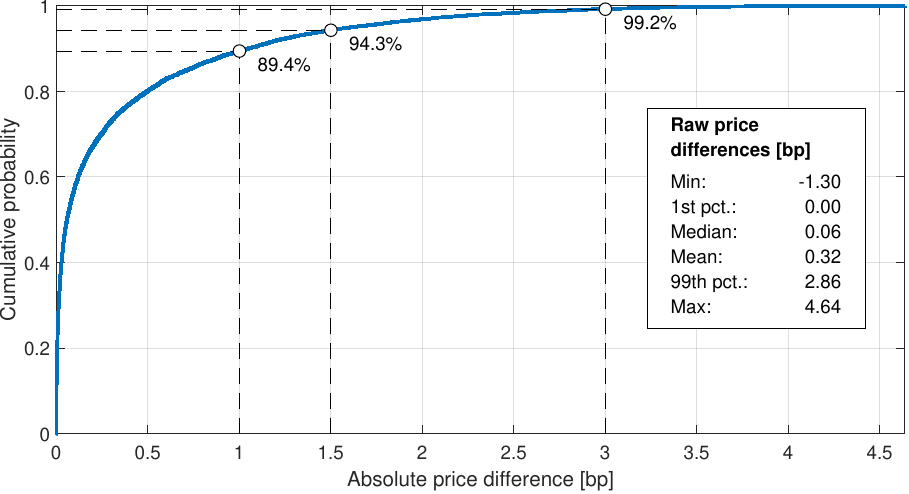}
    \caption{\textbf{Empirical cumulative distribution function of absolute price differences -- European swaptions.} The plot includes markers for the thresholds at \qty{1}{\bp}, \qty{1.5}{\bp} and \qty{3}{\bp} as well as the corresponding percentiles. In addition, a table with basic statistics for the raw price differences is provided.}
    \label{fig:basket_ecdf}
\end{figure}

 Turning to the individual results, we first consider the differences observed for the European swaptions. For transparency, we note that in a handful of cases the upper bound is slightly lower than the benchmark present value. We attribute this to numerical noise and do not address it separately. Figure~\ref{fig:basket_ecdf} shows the empirical cumulative distribution function (ECDF) of absolute deltas. With a median of \qty{0.06}{\bp} and a mean of \qty{0.32}{\bp}, the results reveal that the differences are highly concentrated around zero. In particular, the ECDF shows a limited proportion of observations beyond \qty{1}{\bp}. At \qty{1.5}{\bp}, roughly \qty{94.3}{\percent} of the absolute price differences are smaller, rising to \qty{99.2}{\percent} at \qty{3}{\bp}. 
\begin{table}[tbp]
\centering
\begin{threeparttable}
\caption{\textbf{Effect of naive basket replication on the pricing consistency of amortizing Bermudan swaptions.}}
\label{tab:bermudan_basket_hagav_vs_naive}
\begin{tabular}{lrrrrrrrrrrr}
\toprule
\multirow{2}{*}{Tenor}
& \multicolumn{2}{c}{Calibration}
& \multicolumn{3}{c}{Tail}
& \multicolumn{3}{c}{$p_{ <\cdot}$  }
& \multicolumn{3}{c}{$\overline{\vert\Delta\vert}_{\ge\cdot}$}\\
\cmidrule(lr){2-3}
\cmidrule(lr){4-6}
\cmidrule(lr){7-9}
\cmidrule(lr){10-12}
& $\overline{\Delta t}$& $\overline{\vert\Delta^{\mathrm{m}}\sigma\vert}$ & $P_{95}$ & $P_{99}$ & Max & \qty{1.0}{\bp} & \qty{1.5}{\bp} & \qty{3.0}{\bp} & \qty{1.0}{\bp} & \qty{1.5}{\bp} & \qty{3.0}{\bp}  \\
\midrule
0--5 & 0.52 & 0.01 & 0.01 & 0.01 & 0.01 & 100.0 & 100.0 & 100.0 & - & - & - \\
5--10 & 0.60 & 0.04 & 0.09 & 0.15 & 0.26 & 100.0 & 100.0 & 100.0 & - & - & - \\
10--15 & 0.66 & 0.06 & 0.35 & 0.61 & 0.61 & 100.0 & 100.0 & 100.0 & - & - & - \\
15--20 & 0.69 & 0.11 & 0.93 & 1.38 & 1.52 & 95.8 & 99.4 & 100.0 & 1.22 & 1.52 & - \\
20--25 & 0.71 & 0.21 & 1.98 & 2.59 & 2.66 & 78.5 & 89.5 & 100.0 & 1.59 & 1.97 & - \\
25--30 & 0.73 & 0.22 & 2.69 & 3.30 & 3.47 & 69.6 & 79.0 & 98.9 & 1.90 & 2.22 & 3.45\\
\addlinespace
All & 0.70 & 0.11 & 1.57 & 2.67 & 3.47 & 89.7 & 94.1 & 99.8 & 1.73 & 2.12 & 3.45\\
\bottomrule
\end{tabular}
\begin{tablenotes}[online,flushleft]
\footnotesize
\item[]\hspace*{-\labelsep}\textit{Notes:} $\overline{\Delta t}$ denotes the relative reduction in mean calibration runtime compared to the \enquote{Basket Hagan} approach, reported as a fraction; $\overline{\vert\Delta^\mathrm{m}\sigma\vert}$ is the mean maximum absolute difference of the corresponding volatility steps, expressed in basis points; $p_{ <\cdot}$ stands for the share of instruments with absolute deltas less than a given threshold, expressed in percentages; the final three columns report the mean absolute difference conditional on $|\Delta|$ being greater than or equal to a given threshold, expressed in basis points. All valuation results are presented in basis points relative to the initial notional of the swaption. 
\end{tablenotes}
\end{threeparttable}
\end{table}%
\par
\begin{figure}[tbp]
    \centering
    \includegraphics[width=\textwidth]{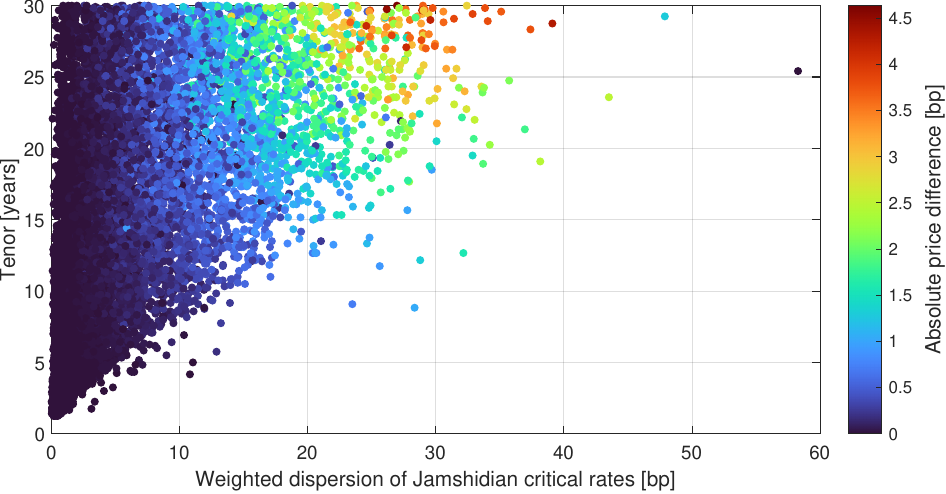}
    \caption{\textbf{Distribution of absolute price differences -- European swaptions.} The scatter plot shows the relationship between the absolute price difference, the tenor of the underlying swap, and dispersion of the Jamshidian critical rates across the basket instruments.}
    \label{fig:basket_scatter_plot_tenor_dispersion}
\end{figure}

 To better assess the tail behavior, we next examine a scatter plot of the delta distribution presented in Figure~\ref{fig:basket_scatter_plot_tenor_dispersion}. It highlights the relationship between the magnitude of the deviations, the tenor of the underlying swap and the dispersion of critical rates from the Jamshidian decomposition. The latter metric aims to indicate whether some basket instruments are in-the-money and others are out-of-the-money at the same time. Under those conditions, the unconditional exercise of every option in the basket is potentially suboptimal and the present value computed by direct summation is expected to be less accurate. We define the dispersion as the standard deviation of the critical rates around their weighted mean. To concentrate on economically material basket instruments, we set the weights to the annuities of the underlying swaps. 
\par Several observations emerge from the figure. First, we note that larger deviations (e.g. $>\qty{1.5}{\bp}$) appear only for longer tenors ($>10$Y). Another notable feature is the distribution of the deltas across the dispersion dimension. Although the observed dispersion of the critical rates ranges from \qtyrange{0}{60}{\bp}, over \qty{75}{\percent} of the observations remain below \qty{10}{\bp}. In this region, only roughly 20 out of over 7,500 swaptions exhibit a difference greater than \qty{1}{\bp}. Moreover, the maximum difference remains below \qty{1.5}{\bp}. An additional pattern becomes apparent when we consider the deltas conditional on a fixed dispersion level. For a given dispersion, longer tenors coincide with higher deviations. Since longer tenors tend to have larger annuities, this reinforces the argument that the observed absolute differences of up to \qty{5}{\bp} could have limited practical relevance for valuation purposes.
\par Finally, we consider the empirical results for Bermudan swaptions presented in Table~\ref{tab:bermudan_basket_hagav_vs_naive}. Based on the findings obtained so far, the metrics are grouped into categories and reported across successive tenor intervals for better readability. Regarding the calibration procedure, we report the relative reduction in mean calibration runtime compared to the \enquote{Basket Hagan} approach $\overline{\Delta t}$, expressed as a fraction, and the mean maximum absolute difference between corresponding volatility steps $\overline{\vert\Delta^{\mathrm{m}}\sigma\vert}$, expressed in basis points. For the pricing category, we provide the following statistics expressed in basis points: the maximum deviation, tail quantiles, and the mean absolute delta for differences greater than or equal to the predefined thresholds~$\overline{\vert\Delta\vert}_{\ge\cdot}$. Additionally, we present the share of absolute deltas less than those thresholds $p_{ <\cdot}$, expressed as a percentage.
\par We begin with the calibration results, which indicate that using the upper bounds in the calibration process does not materially affect the resulting volatility function. The deviations increase with tenor and do not exceed \qty{0.3}{\bp} on average. A closer inspection of the individual calibration runs shows that only 8 out of 1,000 exhibit a difference greater than \qty{1}{\bp}. All correspond to swaptions with at least one of the following parameter values: tenor over 25 years, first exercise under six months and distance to the ATM point of over \qty{250}{\bp}. The mean runtime reduction of roughly \qty{70}{\percent} demonstrates the computational benefit, especially for longer tenors. 
\par Moving to the pricing block, we first note that the results are consistent with prior findings for European swaptions. Comparing the ECDFs of the deviations for Bermudan and European options reveals consistent and partially improved results. This could be attributed to the smoothing effect introduced by the transition from a single exercise date to a sequence of dates distributed over the tenor of the underlying swaps. It is also worth noting that deviations greater than \qty{1.5}{\bp} occur only for tenors longer than 15 years.
\par Taken together, these findings indicate that basing the calibration procedure on the upper bound, rather than on the synthetic prices computed with an intermediate model, does not materially affect the outcome. This holds for both the volatility function and the pricing consistency. From an implementation perspective, omitting the repeated mean reversion calibration substantially reduces computational time, especially for swaptions with tenors exceeding 10 years. 
\subsection{Cross-comparison of calibration sets}
\par The final experiment investigates whether increased sophistication of the calibration set leads to different valuation outcomes. We specifically aim to establish to what extent different calibration sets constructed to address the amortization schedule of the instrument lead to consistent pricing results. With this objective in mind, we compare prices of amortizing Bermudan swaptions computed using all calibration sets presented in the methodology section. Recall that the methods introduced earlier differ in theoretical complexity, computational costs, market data requirements, and especially notional sensitivity.
\par We analyze absolute price differences for a set of \num{1000} randomly generated swaptions with tenors ranging from 2 to 30 years and a quarterly or an annual exercise frequency chosen at random. As with previous experiments, we divide the products into 100 subsets, each evaluated on a randomly assigned date. 
\begin{table}[t]
\small
\centering
\begin{threeparttable}
\caption{\textbf{Cross-comparison of pricing results based on different calibration sets.}}
\label{tab:cross_comparision_methods}
\begin{tabular*}{\textwidth}{@{\extracolsep{\fill}}lrrrrrrrr}
\toprule
Method & Threshold & GC & GCT & LBA & LBC & RS & BH & BN \\
\midrule
Global Column (GC) & 1.5 & - & 15.4 & 10.5 & 4.1 & 4.5 & 4.2 & 4.2 \\
 & 3 & - & 27.7 & 18.1 & 7.3 & 7.4 & 7.3 & 6.9 \\
 & 10 & - & 53.1 & 43.9 & 21.6 & 20.4 & 20.5 & 20.4 \\
\addlinespace[0.5\defaultaddspace]
Global Co-terminal (GCT) & 1.5 & 22.9 & - & 13.8 & 4.3 & 5.1 & 4.5 & 4.6 \\
 & 3 & 26.4 & - & 25.6 & 8.3 & 10.0 & 9.3 & 9.5 \\
 & 10 & 37.5 & - & 58.3 & 23.5 & 25.0 & 25.8 & 25.9 \\
\addlinespace[0.5\defaultaddspace]
Local Bullet ATM (LBA) & 1.5 & 19.2 & 17.3 & - & 6.1 & 3.7 & 4.2 & 4.6 \\
 & 3 & 20.8 & 19.6 & - & 12.5 & 9.2 & 9.5 & 9.8 \\
 & 10 & 27.5 & 30.2 & - & 33.4 & 29.6 & 30.0 & 30.3 \\
\addlinespace[0.5\defaultaddspace]
Local Bullet Core (LBC) & 1.5 & 33.3 & 39.3 & 31.7 & - & 49.6 & 44.3 & 44.9 \\
 & 3 & 34.4 & 40.9 & 33.8 & - & 63.7 & 58.7 & 58.5 \\
 & 10 & 39.5 & 47.9 & 42.5 & - & 86.1 & 86.4 & 86.9 \\
\addlinespace[0.5\defaultaddspace]
Rep. Swaption (RS) & 1.5 & 33.2 & 36.9 & 30.3 & 8.6 & - & 71.5 & 68.5 \\
 & 3 & 34.1 & 38.8 & 32.0 & 11.1 & - & 85.2 & 83.1 \\
 & 10 & 38.6 & 45.4 & 39.4 & 19.9 & - & 98.8 & 98.2 \\
\addlinespace[0.5\defaultaddspace]
Basket Hagan (BH) & 1.5 & 33.3 & 37.2 & 31.0 & 8.0 & 3.9 & - & 95.4 \\
 & 3 & 34.4 & 39.0 & 32.7 & 10.0 & 5.6 & - & 99.2 \\
 & 10 & 39.0 & 46.3 & 40.4 & 18.9 & 11.7 & - & 100.0 \\
\addlinespace[0.5\defaultaddspace]
Basket Naive (BN) & 1.5 & 33.3 & 37.3 & 31.1 & 7.7 & 4.3 & 2.2 & - \\
 & 3 & 34.2 & 39.2 & 32.8 & 9.6 & 6.0 & 3.4 & - \\
 & 10 & 38.9 & 46.5 & 40.5 & 18.4 & 11.9 & - & -\\
\bottomrule
\end{tabular*}
\begin{tablenotes}[online,flushleft]
\footnotesize
\item[]\hspace*{-\labelsep}\textit{Notes:} The table presents a pairwise comparison of valuation results across the different calibration sets introduced in the paper. The upper triangular part of the matrix shows the level of agreement between the methods, defined as the share of valuation runs with an absolute price difference less than a given threshold, expressed as a percentage. The lower triangular part contains the corresponding mean absolute difference conditional on the deviation being greater than or equal to the threshold, expressed in basis points. All metrics are reported for three threshold levels: \qty{1.5}{\bp}, \qty{3}{\bp}, and \qty{10}{\bp}.
\end{tablenotes}
\end{threeparttable}
\end{table}
\par The results reveal several notable patterns. First, as anticipated on theoretical grounds, the choice of the calibration set materially affects the pricing outcome. Even for a relatively large tolerance of \qty{10}{\bp}, some approaches agree in less than \qty{30}{\percent} of cases. Second, and somewhat surprisingly, local calibration without access to smile market data tends to align more closely with simpler global approaches than with other product-dependent methods. Finally, approaches that address the amortization of the underlying swap show a considerably higher degree of alignment. The local calibration using core bullet swaptions serves as a middle ground between global and amortization-sensitive calibration sets. 
\par The detailed results reported in Table~\ref{tab:cross_comparision_methods} provide further insights into these patterns. To begin with, we note that the overall level of agreement varies considerably across methods and thresholds. For the intermediate tolerance of \qty{3}{\bp}, the share of differences less than the threshold ranges from merely \qty{6.9}{\percent} to as much as \qty{99.2}{\percent}. The weakest alignment is observed between global and amortization-sensitive calibration sets. This is also visible in the conditional mean discrepancies ranging from \qtyrange{33.2}{46.5}{\bp}. Consistent with the findings of our previous experiment, the comparison of both basket procedures sits at the opposite end of the scale. Interestingly, the representative swaption approach agrees with both basket approaches in over \qty{83}{\percent} and \qty{98}{\percent} of cases for the \qty{3}{\bp} and \qty{10}{\bp} thresholds, respectively. This demonstrates that methods accounting for the amortization, although exhibiting some noticeable differences, are largely consistent with respect to the overall price level.
\par Based on the discussions so far, the methods can be grouped into two clusters. The first comprises both global methods and the \enquote{Local Bullet ATM} calibration set. The second includes the \enquote{Local Bullet Core} as well as the remaining amortization-sensitive approaches. Although methods within those clusters tend to agree more closely, the pairwise alignment levels still vary substantially. It is noteworthy that for roughly 3 out of 4 swaptions, the lowest present value is produced by the first group. The \enquote{Global Co-terminal} approach generates the minimum price in roughly 35 out of 100 instances. By contrast, the latter cluster produces the highest price in approximately \qty{60}{\percent} of cases.
\par From a practical perspective, the results clearly support the relevance of the calibration set in the valuation of amortizing Bermudan swaptions. Although the experiment cannot determine which method produces the most accurate present values, the fact that amortization-sensitive calibration sets show a substantially higher degree of alignment indicates strong consensus on the overall price level. 
%
%
%
%
\section{Conclusions}\label{sec:conclusions}
\par This paper has investigated to what extent the choice of the calibration set affects the pricing of amortizing Bermudan swaptions under the Hull--White One-Factor model. The approaches considered differ in terms of the required market data, computational costs, sophistication and ability to capture non-standard structures. Specifically, using a multi-year historical dataset of EUR-denominated interest rate curves and a volatility cube, we systematically investigated the resulting pricing differences for a broad spectrum of products. Furthermore, we examined a modification of the notional-sensitive basket method, designed to mitigate computational limitations resulting from nested calibration runs without materially changing the resulting present values.  
\par The numerical results provide evidence that the construction of the calibration set can materially influence the valuation of amortizing Bermudan swaptions. More specifically, the main findings point to two different aspects. On the one hand, complete coverage of the core European swaptions is not a necessary condition to derive consistent prices. After reaching sufficient coverage of the Bermudan exercises, the results indicate a high degree of price consistency and overall stability of the calibrated model parameters. In the considered setting, the saturation point is reached at 15 instruments with only marginal improvements in the examined metrics beyond this threshold. On the other hand, the data reveal that the investigated methods form two clusters depending on whether the smile data from the volatility cube are incorporated. Within each cluster, we observe higher pairwise levels of agreement between prices, although substantial variation remains. In light of these results, two points deserve attention. First, since the \enquote{Local Bullet ATM} method tends to align more closely with global approaches, the results suggest that the product-dependent choice of instruments has less influence on the valuation than the incorporation of smile data. Second, notional-sensitive methods display the highest level of agreement, indicating consensus on the overall price level for amortizing swaptions. In particular, the upper bounds of the prices of amortizing European swaptions are tight enough for the \enquote{Basket Naive} approach to produce meaningful valuations for the corresponding Bermudan swaption.
\par From a practical perspective, these findings suggest that reducing computational complexity, whether by restricting the cardinality of the calibration set or by relying on upper bounds as target prices during basket calibration, can lead to a substantially more efficient pricing procedure without materially affecting the corresponding valuations. This provides the simple answer to the question in the title: substantially less than one might expect. All results presented above should be interpreted subject to the limitation that the present study cannot determine whether any single calibration method is superior in terms of pricing accuracy. The prices of Bermudan swaptions, especially those with an amortizing notional schedule, are generally not observed in the market, preventing a systematic assessment of pricing accuracy. Future research could focus on identifying sufficient conditions under which the upper bound approximations are tight enough to support the calibration procedure. Finally, the analysis may be extended to less systematic amortization schedules. 
%
%
%
%
%
\section*{Acknowledgements}
I would like to thank Prof. Laitenberger (MLU Halle-Wittenberg) for supporting my doctoral research, Prof. Jäger-Ambrożewicz (HTW Berlin) for guidance and constructive feedback, and Man Kit Tsui for carefully reading the manuscript and for valuable discussions throughout this work. Any remaining errors are my own.  
%
%
%
%
\printbibliography
%
%
%
\clearpage
\appendix
\section{Hull--White One-Factor Formulas}\label{app:hw1f_formulas}
\par This appendix provides a detailed derivation of the central results for the Hull--White One-Factor model. Our goal is twofold. First, the following discussion serves as a natural extension of the statements briefly introduced in the methodology section, making the mathematical foundation around the HW1F model self-contained in this paper. Second, showing the sometimes wearisome calculations contributes to the understanding of the mathematical structure of the short-rate models. In what follows, we concentrate on the following aspects: demonstrating that the model has an affine term structure, choosing $\vartheta$ to reproduce the initial term structure, deriving the zero-coupon bond formula $P(t,T)$, and pricing the zero-coupon bond option $\ZBO$. 
\par \textbf{Affine term structure} \newline A short-rate model has an affine term structure\footnote{The representation becomes \enquote{visually} affine for continuously compounded zero-rates.} if the price of the zero-coupon bond can be expressed as 
\begin{equation}\label{eq:general_affine_term_structure}
P(t,T) = \exp ( A(t,T) - r(t)B(t,T) )    
\end{equation}
where $A(t,T)$ and $B(t,T)$ are deterministic functions. At the same time, the general pricing equation for a zero-coupon bond gives
\begin{equation*}
P(t,T)  = \EE^Q \left[ \exp \left( -\int^T_t r(u) \dd u  \right) \given \iF_t   \right].
\end{equation*}
To calculate the expectation and verify whether \eqref{eq:general_affine_term_structure} holds, we first need to determine the distribution of $\int_t^T r(u)\dd u$ under $Q$ conditional on $\iF_t$.
\par To this end, we start by solving the stochastic differential equation describing the evolution of the short rate from \eqref{eq:hw1f_dynamic}. First, we define an auxiliary process $X(t) \coloneqq e^{\iA (0,t)} r(t)$ with $ \iA(t, T) \coloneqq \int^T_t a(u)\dd u$. Applying It\^{o}'s product rule yields 
\begin{align*}
\dd X(t) & = e^{\iA (0,t)} \dd r(t) + r(t)\dd (e^{\iA (0,t)}) +\dd \langle e^{\iA (0,\cdot)} , r(\cdot) \rangle_t \\
      & = e^{\iA (0,t)} \left\{ \left[ \vartheta (t) -a(t)r(t) \right]\dd t+ \sigma (t) \dd W (t) \right\} + \frac{\partial}{\partial t} \left[ e^{\iA (0,t)}\right]r(t)\dd t \\
      & = e^{\iA (0,t)} \vartheta (t) \dd t +e^{\iA (0,t)} \sigma(t) \dd W(t),
\end{align*}
where the quadratic covariation vanishes since $e^{\iA(0,t)}$ is a deterministic function of finite variation and the Leibniz integral rule was applied to evaluate $\frac{\partial}{\partial t} e^{ \iA(0,t)}$. Substituting back into the auxiliary process and solving for $r(t)$ leads to the following integral representation for $0\le s \le t$ 
\begin{equation}\label{eq:hw1f_sde_solution}
r(t) = r(s)e^{-\iA (s,t)}+ \int_s^t e^{-\iA (u,t)} \vartheta (u) \dd u + \int_s^t e^{-\iA (u,t)}\sigma (u)\dd W(u).    
\end{equation}
Note that due to the Gaussianity of the stochastic integral, $r(t)$ is normally distributed under $Q$ conditional on $\iF_s$ with mean $M(s,t)$ and variance $V(s,t)$ given by
\begin{equation}\label{eq:hw1f_mean_variance_r}
M(s,t) = r(s) e^{-\iA (s,t)} + \int_s^t e^{-\iA (u,t)} \vartheta (u) \dd u \quad\text{and} \quad V(s,t) = \int_s^t e^{-2\iA (u,t)}  \sigma^2(u) \dd u.    
\end{equation}
\par Next, we extend the derivation for the constant-parameter case presented in \cite[pp.~169--171]{brigo2006InterestRateModels} to determine the distribution of $\int_t^Tr(u)\dd u$. Given the above result for $r(t)$, we integrate the short rate over $[t,T]$, which yields 
\begin{align}
\int_t^T r(u)\dd u
&= \int_t^T \left[
    r(t)e^{-\iA(t,u)}
    + \int_t^u e^{-\iA(s,u)}\vartheta(s)\dd s
    \right. \nonumber\\
&\qquad\left.
    + \int_t^u e^{-\iA(s,u)}\sigma(s)\dd W(s)
    \right]\dd u \nonumber\\
&= r(t)\int_t^T e^{-\iA(t,u)}\dd u
    + \int_t^T\int_t^u e^{-\iA(s,u)}
      \vartheta(s)\dd s\dd u \nonumber\\
&\qquad
    + \int_t^T\int_t^u e^{-\iA(s,u)}
      \sigma(s)\dd W(s)\dd u \nonumber\\
&= r(t)\int_t^T e^{-\iA(t,u)}\dd u
    + \int_t^T\vartheta(s)
      \int_s^T e^{-\iA(s,u)}\dd u\dd s \nonumber\\
&\qquad
    + \int_t^T\sigma(s)
      \int_s^T e^{-\iA(s,u)}\dd u\dd W(s).
\label{eq:hw1f_solution_integral_short_rate}
\end{align}
Observe that in the last step, the deterministic and stochastic Fubini theorems were applied to change the order of integration. In particular, we use the following equivalence of sets for $0\le t \le T$
\begin{equation*}
\left\{ (u,s)\in \RR^2 \Big|{} t \le u \le T, t \le s \le u \right \} = \left\{ (u,s)\in \RR^2 \Big|{} t \le s \le T , s \le u \le T \right\}.  
\end{equation*}
An inspection of the right-hand side of \eqref{eq:hw1f_solution_integral_short_rate} yields that $ -\int_t^T r(u)\dd u$ is normally distributed under $Q$ conditional on $\iF_t$ with expected value $\hat{M}(t,T)$ and variance $\hat{V}(t,T)$ given by  
\begin{align}
  \hat{M}(t,T) &=  -r(t) \int^T_t e^{-\iA (t,u)} \dd u - \int^T_t \vartheta (s) \int_s^T e^{-\iA (s,u)} \dd u \dd s  \nonumber\\ 
\hat{V}(t,T) & = \int^T_t \sigma^2 (s) \left( \int_s^T e^{-\iA (s,u)}\dd u \right)^2 \dd s. \label{eq:hw1f_mean_variance_of_int_r_tT}   
\end{align}
\par Finally, the relationship between normal and lognormal random variables implies 
\begin{equation}\label{eq:hw1f_zero_coupond_bond_expectation_result}
P(t,T) = \EE^Q \left[ \exp\left(  -\int_t^T r(u) \dd u  \right) \given \iF_t\right] = \exp \left(\hat{M}(t,T) + \frac{1}{2}\hat{V}(t,T) \right).
\end{equation}
Matching terms in \eqref{eq:general_affine_term_structure} and \eqref{eq:hw1f_zero_coupond_bond_expectation_result} gives
\begin{align}
B(t,T) & = \int^T_t e^{-\iA (t,u)}\dd u \nonumber \\
A(t,T) & = -\int_t^T \vartheta (s) B(s,T)\dd s + \frac{1}{2} \int_t^T \sigma^2 (s) B^2(s,T) \dd s, \label{eq:hw1f_a_b_functions}
\end{align}
demonstrating that the Hull--White One-Factor model has an affine term structure.
\par \textbf{Initial term structure fitting} \newline In the next step, we determine the drift term $\vartheta (t)$ in such a way that the previously derived model prices of zero-coupon bonds $P(0, T)$ coincide with market observations given by the discount curve
\begin{equation*}
\{ T \mapsto P^\mathrm{M}(0,T), T>0 \}.
\end{equation*} 
To achieve this, we follow the approach presented in \cite[pp.~407--412]{andersen2010InterestRateModeling} and impose  
\begin{align*}
P^\mathrm{M} (0, T) \overset{!}{=} P(0, T).
\end{align*}
Taking logarithms, differentiating with respect to the maturity $T$ on both sides, multiplying both sides by $-1$ and substituting the affine representation in \eqref{eq:general_affine_term_structure} results in  
\begin{align}
-\frac{\partial}{\partial T} \ln P^\mathrm{M} (0,T) & \overset{!}{=} -\frac{\partial}{\partial T} \ln P (0,T) \nonumber \\
f^\mathrm{M}(0,T) & \overset{!}{=} r(0) \frac{\partial}{\partial T} B(0,T)  - \frac{\partial}{\partial T} A(0,T),\label{eq:hw1f_drift_term_equation}
\end{align}
where the definition of the instantaneous forward rate $f^\mathrm{M}(t,T)$ was used in the second equation. Since $A(t,T)$ is defined in terms of $\vartheta (t)$, a connection between the model-induced zero-coupon bond prices and the instantaneous forward curve has been established. 
\par Substituting the previously derived expressions for $A(0,T)$ and $B(0,T)$ requires differentiating several integral expressions, which we evaluate using the Leibniz integral rule. First, for the loading function $B(0,T)$, we get  
\begin{equation}\label{eq:derivative_loading_function_dT_BtT}
\frac{\partial}{\partial T} B(0,T) = e^{-\iA (0,T)}.
\end{equation}
Next, the first integral arising from $A(0,T)$ yields
\begin{align*}
\frac{\partial}{\partial T} \left( -\int_0^T  \vartheta (s) B(s,T)\dd s \right) & = - \int_0^T \frac{\partial}{\partial T} \left( \vartheta(s) B(s,T) \right) \dd s \\
&\qquad -\vartheta(T)B(T,T) \frac{\partial}{\partial T}(T) + \vartheta(0)B(0,T)\frac{\partial}{\partial T}(0) \\
& = -\int_0^T \vartheta (s) e^{-\iA(s,T)}\dd s,
\end{align*}
where $B(T,T)=0$ and $\frac{\partial}{\partial T} (0)=0$ were used. Using the simplified expression from \eqref{eq:hw1f_mean_variance_of_int_r_tT}  
\begin{align*}
\hat{V}(0,T) = \int_0^T \sigma^2 (s)B^2(s,T) \dd s, 
\end{align*}
and substituting into \eqref{eq:hw1f_drift_term_equation} leads to
\begin{equation}\label{eq:hw1f_drift_term_equation_after_substitution}
f^\mathrm{M}(0,T) = r(0)e^{-\iA (0,T)} + \int_0^T \vartheta (s) e^{-\iA(s,T)}\dd s - \frac{1}{2} \frac{\partial}{\partial T} \hat{V}(0,T).
\end{equation}
The initial term structure fitting problem thus reduces to determining the drift term $\vartheta (t)$ by solving the above equation. 
\par To achieve this, we follow the approach presented in \cite[pp.~383--386]{bjork2009ArbitrageTheoryContinuous} and interpret the right-hand side of \eqref{eq:hw1f_drift_term_equation_after_substitution} as $f^\mathrm{M}(0,T) =x(T)-g(T)$ with $x(0) =r(0)$ and
\begin{align}
x(T) & \coloneqq r(0)e^{-\iA (0,T)} + \int_0^T \vartheta (s) e^{-\iA(s,T)}\dd s \nonumber \\
g(T) & \coloneqq \frac{1}{2} \frac{\partial}{\partial T} \hat{V}(0,T). \label{eq:hw1f_drift_derivation_aux_functions}
\end{align}
Applying the Leibniz integral rule to calculate the derivative of $x(T)$ yields
\begin{align*}
\frac{\partial}{\partial T} x(T) &= r(0)\frac{\partial }{\partial T} \left( e^{-\iA (0,T)} \right) + \frac{\partial}{\partial T} \left( \int_0^T \vartheta (s) e^{-\iA (s,T)} \dd s\right) \\
& = -a(T)e^{-\iA (0,T)} r(0) + \int_0^T \vartheta (s) \frac{\partial}{\partial T} \left( e^{-\iA (s,T)} \right)\dd s  \\
& \qquad + \vartheta (T) e^{-\iA(T,T)}\frac{\partial}{\partial T} (T) - \vartheta (0) e^{-\iA(0,T)}\frac{\partial}{\partial T} (0) \\
& = -a(T)e^{-\iA (0,T)} r(0) -a(T) \int_0^T \vartheta (s) e^{-\iA (s,T)} \dd s + \vartheta(T) \\
& = -a(T)\left( e^{-\iA (0,T)} r(0) + \int_0^T \vartheta (s)  e^{-\iA (s,T)}\dd s \right) + \vartheta (T) \\
& = -a(T)x(T) + \vartheta(T)
\end{align*}
and therefore
\begin{align*}
\vartheta(T) & = a(T)x(T) + \frac{\partial}{\partial T} x(T) \\
             & = a(T) \left( f^\mathrm{M}(0,T) + g(T)\right) + \frac{\partial}{\partial T} \left( f^\mathrm{M}(0,T) + g(T) \right) \\
             & = a(T)f^\mathrm{M}(0,T)+ \frac{\partial}{\partial T} f^\mathrm{M}(0,T) + a(T)g(T)+ \frac{\partial}{\partial T} g(T).
\end{align*} 
Substituting for $g(T)$ leads to 
\begin{align*}
    \vartheta(t) & = a(t)f^\mathrm{M}(0,t) + \frac{\partial}{\partial t} f^\mathrm{M}(0,t) + \frac{1}{2} \left[ a(t) \frac{\partial }{\partial t} \hat{V}(0,t) + \frac{\partial^2}{\partial t^2} \hat{V}(0,t) \right] \\
    & \eqqcolon a(t)f^\mathrm{M}(0,t) + \frac{\partial}{\partial t} f^\mathrm{M}(0,t) +\frac{1}{2}I
\end{align*}
Again, using the Leibniz integral rule to simplify $I$ provides
\begin{align*}
I & =  a(t) \frac{\partial }{\partial t} \hat{V}(0,t) + \frac{\partial^2}{\partial t^2} \hat{V}(0,t) \\ 
& = a(t) \frac{\partial }{\partial t} \int_0^t \sigma^2 (u) B^2(u,t) \dd u +\frac{\partial^2 }{\partial t^2} \int_0^t \sigma^2 (u) B^2(u,t) \dd u \\
& = \int_0^t \sigma^2 (u) \left[ a(t) \frac{\partial }{\partial t} B^2 (u,t) + \frac{\partial^2 }{\partial t^2} B^2(u,t) \right] \dd u \\
& = \int_0^t \sigma^2 (u) \left[ a(t) 2 B(u,t) \frac{\partial }{\partial t} B (u,t) + \frac{\partial }{\partial t} \left[2B(u,t) \frac{\partial}{\partial t}B(u,t) \right] \right] \dd u \\
& = 2 \int_0^t \sigma^2 (u) \left[ B(u,t)a(t)e^{-\iA (u,t)} +e^{-2\iA (u,t)} - B(u,t)a(t)e^{-\iA (u,t)} \right] \dd u \\
& = 2 \int_0^t \sigma^2 (u) e^{-2\iA (u,t)} \dd u \\
& = 2 V(0, t),
\end{align*}
where \eqref{eq:derivative_loading_function_dT_BtT} and \eqref{eq:hw1f_mean_variance_r} were used. Finally, combining the above results yields the form of the drift rate consistent with the initial term structure
\begin{equation}\label{eq:hw1f_drift_rate_final_derivation_result}
\vartheta(t)  = a(t)f^\mathrm{M}(0,t) + \frac{\partial}{\partial t} f^\mathrm{M}(0,t) + V(0,t).
\end{equation}
\par \textbf{Zero-coupon bond} \newline To derive an explicit pricing formula for the zero-coupon bond $P(t,T)$, we need to insert the final form of $\vartheta(t)$ into the risk-neutral expectation from \eqref{eq:hw1f_zero_coupond_bond_expectation_result}. Substituting the expression for $\vartheta(t)$ from  \eqref{eq:hw1f_drift_rate_final_derivation_result} into $\hat{M}(t,T)$ yields 
\begin{align*}
\hat{M}(t,T) & = -r(t)B(t,T) -\int_t^T a(s)f^\mathrm{M}(0,s)B(s,T)\dd s-\int_t^T \frac{\partial}{\partial s}\left[f^\mathrm{M}(0,s)\right]  B(s,T)\dd s \\ 
& \qquad - \frac{1}{2}\int_t^T  \left( a(s)\frac{\partial}{\partial s} \left[\hat{V}(0,s)\right]B(s,T) +\frac{\partial^2}{\partial s^2 }\left[\hat{V}(0,s)\right] B(s,T)\right)  \dd s \\
 & =: -r(t)B(t,T) -I_1 -I_2 - I_3.
\end{align*}
The goal is to evaluate the integrals $I_1$ to $I_3$ and simplify the resulting terms.
\par Starting with $I_2$, integration by parts leads to
\begin{align*}
\int_t^T\frac{\partial}{\partial s} \left[ f^\mathrm{M}(0,s)\right] B(s,T)\dd s & = \left[ f^\mathrm{M}(0,s)B(s,T) \right]^T_t - \int_t^T f^\mathrm{M}(0,s) \frac{\partial}{\partial s} B(s,T) \dd s \\
& = f^\mathrm{M}(0,T)B(T,T) -f^\mathrm{M}(0,t)B(t,T)  \\
& \qquad - \int_t^T f^\mathrm{M}(0,s) \frac{\partial}{\partial s} B(s,T) \dd s \\
& = -f^\mathrm{M}(0,t)B(t,T) - \int_t^T f^\mathrm{M}(0,s) \frac{\partial}{\partial s} B(s,T) \dd s.
\end{align*}
Overall, we obtain the following relationship between $I_2$ and $I_1$
\begin{align*}
I_1 + I_2 & = -f^\mathrm{M}(0,t) B(t,T) + \int_t^T a(s)f^\mathrm{M}(0,s)B(s,T) \dd s  -\int_t^T f^\mathrm{M}(0,s) \frac{\partial}{\partial s} B(s,T) \dd s \\
& = -f^\mathrm{M}(0,t) B(t,T) + \int_t^T f^\mathrm{M}(0,s) \left[ a(s)B(s,T) - \frac{\partial}{\partial s}B(s,T) \right]\dd s  \\
&  =  -f^\mathrm{M}(0,t) B(t,T) +\int^T_t f^\mathrm{M}(0,s)\dd s,
\end{align*}
where the property $\frac{\partial}{\partial t}B(t,T) = a(t) B(t,T) -1$ was used to obtain the last line. 
\par Next, we split $I_3$ into two integrals
\begin{align*}
I_3 & = \frac{1}{2}\int_t^T  \left( a(s)\frac{\partial}{\partial s} \left[\hat{V} (0,s)\right]B(s,T) +\frac{\partial^2}{\partial s^2 }\left[\hat{V}(0,s)\right] B(s,T)\right) \dd s \\
& = \frac{1}{2}\int_t^T a(s)\frac{\partial}{\partial s} \left[\hat{V}(0,s)\right]B(s,T)\dd s + \frac{1}{2} \int_t^T \frac{\partial^2}{\partial s^2 }\left[\hat{V}(0,s)\right] B(s,T) \dd s \\
& =: \frac{1}{2} I_{3|1} +\frac{1}{2} I_{3|2}.
\end{align*} 
Reapplying the previously established identity for $\frac{\partial}{\partial t} B(t,T)$ yields 
\begin{align*}
I_{3|1} & = \int_t^T a(s)\frac{\partial}{\partial s} \left[\hat{V}(0,s)\right]B(s,T)\dd s  \\
& = \int_t^T \left[ 1 + \frac{\partial}{\partial s} B(s,T) \right] \frac{\partial}{\partial s} \hat{V}(0,s)\dd s  \\
& = \int_t^T \frac{\partial}{\partial s}\left[\hat{V}(0,s)\right]\dd s  +  \int_t^T \frac{\partial}{\partial s} \left[ B(s,T) \right] \frac{\partial}{\partial s} \left[\hat{V}(0,s)\right] \dd s \\
& =: \hat{V}(0,T) - \hat{V}(0,t) + I^*
\end{align*}
with $I^*$ being a deterministic integral. Moreover, integration by parts produces  
\begin{align*}
I_{3|2} & = \int_t^T \frac{\partial^2}{\partial s^2 }\left[\hat{V}(0,s)\right] B(s,T) \dd s \\
& = \left[ \frac{\partial}{\partial s }\left[\hat{V}(0,s)\right] B(s,T) \right]_t^T - \int_t^T \frac{\partial}{\partial s} \left[ B(s,T) \right] \frac{\partial}{\partial s} \left[\hat{V}(0,s)\right] \dd s \\
& = \frac{\partial}{\partial T }\left[\hat{V}(0,T)\right]B(T,T) - \frac{\partial}{\partial t }\left[\hat{V}(0,t)\right] B(t,T) - I^* \\
& = - \frac{\partial}{\partial t }\left[\hat{V}(0,t)\right] B(t,T) - I^*.
\end{align*}
Combining the expressions, we get the following result for $I_3$
\begin{align*}
I_3 & = \frac{1}{2} \left[ \hat{V}(0,T)-\hat{V}(0,t)  - \frac{\partial}{\partial t }\left[\hat{V}(0,t)\right] B(t,T) \right] \\
& = \frac{1}{2} \hat{V}(t,T) +\frac{1}{2} \left[ \int_0^t \sigma^2(u)B^2(u,T)\dd u -\hat{V}(0,t) - \frac{\partial}{\partial t }\left[\hat{V}(0,t)\right] B(t,T)   \right] \\
& =: \frac{1}{2}\hat{V}(t,T) + \frac{1}{2}\Delta,
\end{align*}
where the second equality follows from 
\begin{align*}
\hat{V}(0,T) = \hat{V}(t,T) +  \int_0^t \sigma^2(u)B^2(u,T)\dd u.
\end{align*}
\par The expression in $\Delta$ can be simplified by applying the Leibniz integral rule
\begin{align*}
\Delta & =  \int_0^t \sigma^2(u)B^2(u,T)\dd u -\hat{V}(0,t) - \frac{\partial}{\partial t }\left[\hat{V}(0,t)\right] B(t,T) \\
& =   \int_0^t \sigma^2(u)B^2(u,T)\dd u -  \int_0^t \sigma^2(u)B^2(u,t)\dd u - B(t,T) \int_0^t \sigma^2 (u) \frac{\partial}{\partial t } \left[ B^2(u,t)\right]\dd u \\
& = B(t,T) \int_0^t \sigma^2 (u)  \left[  \frac{ B^2(u,T) - B^2(u,t)- \frac{\partial }{\partial t} \left[B^2(u,t) \right]B(t,T) }{B(t,T)}   \right] \dd u \\
& =: B(t,T) \int_0^t \sigma^2 (u) \delta(u) \dd u.
\end{align*}
Moreover, using the following property for $T>S$
\begin{equation}\label{eq:app_property_B_difference}
B(t,T)-B(t,S) = e^{-\iA(t,S)} B(S,T),
\end{equation}
the expression for $\delta (u)$ can be reduced to
\begin{align*}
\delta(u) & =\frac{ B^2(u,T) - B^2(u,t)- \frac{\partial }{\partial t} \left[B^2(u,t) \right]B(t,T) }{B(t,T)} \\
& =  \frac{ (B(u,T) - B(u,t))(B(u,T) + B(u,t))- 2B(u,t)\frac{\partial }{\partial t} \left[B(u,t) \right]B(t,T) }{B(t,T)} \\
& = \frac{ B(t,T)e^{-\iA (u,t)}(B(u,T) + B(u,t))- 2B(u,t)e^{-\iA (u,t)}B(t,T) }{B(t,T)} \\
& = e^{-\iA (u,t)} \left[ B(u,T) - B(u,t)\right] \\
& = B(t,T) e^{-2\iA (u,t)}   
\end{align*}
and therefore
\begin{align*}
\Delta & = B^2(t,T) \int_0^t \sigma^2 (u) e^{-2\iA (u,t) }\dd u \\
	   & = B^2(t,T) V(0,t).
\end{align*}
\par Finally, substituting the above results yields the zero-coupon bond pricing formula
\begin{align*}
P(t,T)	& = \EE^Q \left[ \exp \left( -\int^T_t r(u) \dd u  \right) \given \iF_t   \right] \\
		& = \exp \left( \hat{M}(t,T) + \frac{1}{2}\hat{V}(t,T) \right) \\
		& = \exp \left( -r(t) B(t,T) - (I_1+I_2) -I_3 + \frac{1}{2}\hat{V}(t,T) \right) \\
		& = \exp \Bigg( -r(t) B(t,T) +f^\mathrm{M}(0,t)B(t,T) - \int_t^T f^\mathrm{M}(0,u) \dd u - \frac{1}{2}\Delta \Bigg) \\
		& = \exp \left( -\int_t^T f^\mathrm{M}(0,u) \dd u \right)	\exp \left( f^\mathrm{M} (0,t) B(t,T) -r(t)B(t,T) - \frac{1}{2}\Delta\right)  \\
		& = \exp \left( -\int_0^T f^\mathrm{M}(0,u) \dd u + \int_0^t f^\mathrm{M}(0,u) \dd u \right) \\
			& \qquad \times \exp \left( f^\mathrm{M}(0,t) B(t,T) -r(t) B(t,T) -\frac{1}{2}\Delta\right) \\
		& = \frac{P^\mathrm{M}(0,T)}{P^\mathrm{M}(0,t)} \exp \left( f^\mathrm{M}(0,t) B(t,T) - r(t)B(t,T) - \frac{1}{2}B^2 (t,T)V(0,t) \right) \\
        & = \frac{P^\mathrm{M}(0,T)}{P^\mathrm{M}(0,t)} \exp \left( - x(t)B(t,T) - \frac{1}{2}B^2 (t,T)V(0,t) \right).
\end{align*} 
\par \textbf{Zero-coupon bond option} \newline A European zero-coupon bond option $\ZBO$ with exercise date $T^\mathrm{E}$ is an agreement between two parties that gives the holder the right, but not the obligation, to buy (call option) or sell (put option) a zero-coupon bond with maturity $T^\mathrm{M}>T^\mathrm{E}$ at a predefined strike price $K$.  
\par In what follows, we adopt the approach of \cite[pp.~406--411]{bjork2009ArbitrageTheoryContinuous}. To derive the HW1F price of the zero-coupon bond option $\ZBO_t (T^\mathrm{E}, T^\mathrm{M}, K, \omega)$ at time $t\le T^\mathrm{E}$, we first consider its payoff at $T^\mathrm{E}$
\begin{align*}
\ZBO_{T^\mathrm{E}} (T^\mathrm{E}, T^\mathrm{M}, K, \omega ) = \left[ \omega \left( P(T^\mathrm{E},T^\mathrm{M})-K \right) \right]^+,
\end{align*}
where $\omega = 1$ denotes a call option and $\omega = -1$ a put option. The general risk-neutral pricing formula gives the price at $t\le T^\mathrm{E}$
\begin{align*}
\ZBO_t  (T^\mathrm{E} , T^\mathrm{M}, K, \omega ) \coloneqq \EE^Q \left[ \exp\left(  -\int_t^{T^\mathrm{E}} r(u) \dd u  \right) \left[ \omega \left( P(T^\mathrm{E},T^\mathrm{M})-K \right) \right]^+ \given \iF_t \right].
\end{align*}
To calculate the expectation, we decompose it into two terms
\begin{align}\label{eq:zbo_ansatz}
\ZBO_t (T^\mathrm{E}, T^\mathrm{M}, K, \omega ) & = \EE^Q \left[ \exp\left(  -\int_t^{T^\mathrm{E}} r(u) \dd u  \right)\left( \omega \left( P(T^\mathrm{E},T^\mathrm{M})-K \right) \right)^+ \given \iF_t\right] \nonumber \\ 
& = \omega \EE^Q \left[ \exp\left(  -\int_t^{T^\mathrm{E}} r(u) \dd u  \right)  P(T^\mathrm{E},T^\mathrm{M}) \mathbf{1}_{ \{ \omega \left( P(T^\mathrm{E},T^\mathrm{M})-K \right) \ge 0  \} } \given \iF_t\right] \nonumber\\
& \qquad - \omega K \EE^Q \left[ \exp\left(  -\int_t^{T^\mathrm{E}} r(u) \dd u  \right)   \mathbf{1}_{ \{ \omega \left( P(T^\mathrm{E},T^\mathrm{M})-K \right) \ge 0  \} } \given \iF_t\right] \nonumber\\
& =: \omega E_1 - \omega KE_2,
\end{align}
where $\mathbf{1}_{\{\cdot \}}$ denotes the indicator function. Using the change-of-numéraire technique leads to
\begin{align*}
\ZBO_t (T^\mathrm{E}, T^\mathrm{M}, K,\omega) & = \omega E_1-\omega KE_2  \nonumber \\
& =  \omega P(t,T^\mathrm{M}) \EE ^{T^\mathrm{M}} \left[ \frac{P(T^\mathrm{E},T^\mathrm{M})}{P(T^\mathrm{E},T^\mathrm{M})} \mathbf{1}_{ \{ \omega \left( P(T^\mathrm{E},T^\mathrm{M})-K \right) \ge 0  \} } \given \iF_t \right]  \nonumber \\
	& \qquad - \omega K P(t,T^\mathrm{E}) \EE^{T^\mathrm{E}} \left[ \frac{1}{P(T^\mathrm{E},T^\mathrm{E})} \mathbf{1}_{ \{ \omega \left( P(T^\mathrm{E},T^\mathrm{M})-K \right) \ge 0  \} } \given \iF_t \right]  \nonumber \\
& = \omega P(t,T^\mathrm{M})Q^{T^\mathrm{M}} \left[ \omega \left( P(T^\mathrm{E},T^\mathrm{M})-K \right) \ge 0 \given \iF_t \right]  \nonumber \\
	& \qquad - \omega K P(t,T^\mathrm{E}) Q^{T^\mathrm{E}} \left[ \omega \left( P(T^\mathrm{E},T^\mathrm{M})-K \right) \ge 0  \given \iF_t\right] \nonumber \\
& =: \omega P(t,T^\mathrm{M}) W_1 - \omega K P(t,T^\mathrm{E}) W_2.
\end{align*}
The valuation thus reduces to determining the probabilities $W_1$ and $W_2$.
\par To calculate $W_1$ and $W_2$, we consider the auxiliary process $Z(t) \coloneqq \frac{ P(t,T^\mathrm{M}) }{ P(t,T^\mathrm{E})}$ whose dynamics take the form 
\begin{equation}\label{eq:hw1f_zbo_auxiliary_process_z_evolution}
\dd Z(t) = Z(t) m_Z(t,Z(t)) \dd t + Z(t) \sigma_Z (t) \dd W(t).
\end{equation}
To verify this, we use the general affine term structure formula from \eqref{eq:general_affine_term_structure} and get
\begin{align*}
    Z(t) & = \frac{P(t,T^\mathrm{M})}{P(t,T^\mathrm{E})} = \exp \big( A(t,T^\mathrm{M})-A(t,T^\mathrm{E}) - \left [ B(t,T^\mathrm{M})-B(t,T^\mathrm{E}) \right] r(t) \big) \\
    & \eqqcolon \exp\left( H(t,T^\mathrm{E},T^\mathrm{M},r(t)) \right).
\end{align*}
Given $Z(t) = f(t,r(t))$, applying It\^o's formula yields 
\begin{equation*}
\dd Z(t) = \frac{\partial}{\partial t} f(t,r(t)) \dd t + \frac{\partial}{\partial r} f(t,r(t)) \dd r(t) + \frac{1}{2}\frac{\partial^2}{\partial r^2} f(t,r(t))\dd \langle r(\cdot) \rangle_t.
\end{equation*}
Substitution of the following derivatives and the short-rate dynamics
\begin{align*}
\frac{\partial}{\partial t}f (t,r(t)) & = Z(t) \frac{\partial}{\partial t} H(t,T^\mathrm{E},T^\mathrm{M},r(t)) \\
\frac{\partial}{\partial r}f (t,r(t)) & = -Z(t) \left[ B(t,T^\mathrm{M}) -B(t,T^\mathrm{E}) \right] \\
\frac{\partial^2}{\partial r^2} f(t,r(t)) & = Z(t) \left[  B(t,T^\mathrm{M}) -B(t,T^\mathrm{E}) \right] ^2
\end{align*}
leads to
\begin{equation}\label{eq:hw1f_zbo_auxiliary_process_z_evolution_derived}
\dd Z(t) = Z(t) \left[ \cdots \right] \dd t - Z(t) \sigma (t) \left[ B(t,T^\mathrm{M}) - B(t,T^\mathrm{E}) \right] \dd W(t),
\end{equation}
where the exact form of the drift term is omitted. Matching terms between \eqref{eq:hw1f_zbo_auxiliary_process_z_evolution} and \eqref{eq:hw1f_zbo_auxiliary_process_z_evolution_derived} gives us
\begin{equation*}
\sigma_Z(t) = -\sigma(t) \left[ B(t,T^\mathrm{M}) - B(t,T^\mathrm{E}) \right].
\end{equation*}
\par Note that $Z(t)$ is a $Q^{T^\mathrm{E}}$ martingale and therefore its dynamics under $Q^{T^\mathrm{E}}$ are given by
\begin{equation*}
\dd Z(t) = Z(t) \sigma_Z (t) \dd W^{T^\mathrm{E}}(t).
\end{equation*}
Moreover, the process $Z(\cdot)$ is a time-inhomogeneous geometric Brownian motion under $Q^{T^\mathrm{E}}$ with 
\begin{align*}
Z(T^\mathrm{E}) &  = Z(t) \exp\left( -\frac{1}{2} \int_t^{T^\mathrm{E}} \sigma_Z^2 (u) \dd u +\int_t^{T^\mathrm{E}} \sigma_Z(u) \dd W^{T^\mathrm{E}}(u) \right) \\
& =: \frac{P(t,T^\mathrm{M})}{P(t,T^\mathrm{E})}\exp (X),
\end{align*}
where $X$ is normally distributed under $Q^{T^\mathrm{E}}$ conditional on $\iF_t$ with mean $-\frac{1}{2}\theta$ and variance $\theta$ given by
\begin{equation}\label{eq:hw1g_zbo_intergal_theta}
\theta \coloneqq \int_t^{T^\mathrm{E}} \sigma_Z^2(u)\dd u.
\end{equation}
Overall, this leads to the following expression for the second probability with $\omega =1$
\begin{align*}
\left.W_2\right|_{\omega =1} & = Q^{T^\mathrm{E}} \left( P(T^\mathrm{E},T^\mathrm{M}) \ge K \big| \iF_t \right) \\
    & = Q^{T^\mathrm{E}} \left( \frac{P(T^\mathrm{E},T^\mathrm{M})}{P(T^\mathrm{E},T^\mathrm{E})} \ge \frac{K}{1} \bigg| \iF_t \right)   \\
    & = Q^{T^\mathrm{E}} \left( Z(T^\mathrm{E}) \ge K \big| \iF_t \right) \\
    & = Q^{T^\mathrm{E}} \left( \frac{P(t,T^\mathrm{M})}{P(t,T^\mathrm{E})} \exp (X) \ge K \bigg| \iF_t \right) \\
    & = Q^{T^\mathrm{E}} \left( X \ge \ln \left( \frac{P(t,T^\mathrm{E})K}{P(t,T^\mathrm{M})} \right) \bigg| \iF_t \right) \\
    & = Q^{T^\mathrm{E}} \left( \frac{X+\frac{1}{2}\theta}{\sqrt{\theta}} \ge \frac{\ln\left(\frac{P(t,T^\mathrm{E})K}{P(t,T^\mathrm{M})} \right) + \frac{1}{2} \theta }{\sqrt{\theta}} \bigg|\iF_t \right) \\
    & = 1 - \Phi  \left( \frac{\ln\left(\frac{P(t,T^\mathrm{E})K}{P(t,T^\mathrm{M})} \right) + \frac{1}{2} \theta }{\sqrt{\theta}}   \right) \\
    & = \Phi \left( \frac{\ln\left(\frac{P(t,T^\mathrm{M})}{P(t,T^\mathrm{E})K} \right) - \frac{1}{2} \theta }{\sqrt{\theta}} \right) =: \Phi (d),
\end{align*}
where the last equality follows from the symmetry of the standard normal distribution $\Phi (-x) = 1-\Phi (x)$. An analogous computation for $\omega =-1$ yields 
\begin{align}\label{eq_zbo_final_prob_w2}
\left.W_2\right|_{\omega =-1} & = \Phi ( -d ), \nonumber \\
W_2 & = \Phi (\omega d).
\end{align}
\par An analogous treatment of the remaining probability based on an auxiliary process given by $Y(t) = Z(t)^{-1}$ yields $W_1 = \Phi  \left( \omega \left( d + \sqrt{\theta} \right) \right)$. Note that $Y(t)$ is a martingale under the forward measure $Q^{T^\mathrm{M}}$. Finally, the expression for $\theta$ can be simplified using \eqref{eq:app_property_B_difference}, leading to
\begin{align*}
\theta & = \int_t^{T^\mathrm{E}} \sigma_Z^2 (u) \dd u \\
       & = \int^{T^\mathrm{E}}_t \left[ -\sigma(u) \left( B(u,T^\mathrm{M}) -B(u,T^\mathrm{E}) \right) \right]^2 \dd u \\
       & = \int^{T^\mathrm{E}}_t \sigma^2 (u) e^{-2\iA (u,T^\mathrm{E})}B^2(T^\mathrm{E},T^\mathrm{M})\dd u \\
       & = B^2(T^\mathrm{E},T^\mathrm{M}) V(t,T^\mathrm{E}).
\end{align*}
Combining the above results yields the zero-coupon bond option price 
\begin{equation*}
    \ZBO_t (T^\mathrm{E}, T^\mathrm{M}, K,\omega) = \omega P(t,T^\mathrm{M}) \Phi  \left( \omega \left( d + \sqrt{\theta} \right) \right) - \omega K P(t,T^\mathrm{E}) \Phi (\omega d)
\end{equation*}
with 
\begin{align*}
    d & =  \frac{\ln\left(\frac{P(t,T^\mathrm{M})}{P(t,T^\mathrm{E})K} \right) - \frac{1}{2} \theta }{\sqrt{\theta}} \\
    \theta & = B^2(T^\mathrm{E},T^\mathrm{M}) V(t,T^\mathrm{E}).
\end{align*}
\clearpage
\section{Bachelier model}\label{app:bachelier_model}
\par This appendix provides a detailed derivation of the price of a European swaption on a plain vanilla interest rate swap $\IRS(T^S,T^M, K-S^\mathrm{p}(t), N)$ under the Bachelier model. Recall that the model specifies the dynamics of the par rate of the underlying swap for $u\ge t$ by
\begin{equation}
    \dd S(u) = \sigma_\mathrm{B}\mathrm{d}W^C(u) \quad\text{with} \quad S(t) = S^\mathrm{p}(t)\label{eq:bachelier_model} 
\end{equation}
where $S^\mathrm{p}(t)$ denotes the current par swap rate and $W^C(u)$ the Brownian motion under the annuity measure $Q^C$ associated with the annuity of the underlying swap acting as numéraire. Note that we drop the superscript $\mathrm{p}$ from the par rate $S^\mathrm{p}$ in \eqref{eq:bachelier_model} to ease the notation.
\par Given these dynamics and the price formula for a swap established in \eqref{eq:irs_price_through_swap_rate}, the price of a European swaption with exercise at $T^\mathrm{E}$ and settlement date $T^\mathrm{S}$ on a swap $\IRS(T^S,T^M, K-S(t), N)$ is determined by the following expectation
\begin{equation*}
    \ESWO^\mathrm{B}_t (T^\mathrm{E},T^\mathrm{S},\IRS) = C(t) \EE^C \left[  \left(\omega\left( K-S(T^\mathrm{E}) \right) \right)^+ \given\iF_t \right].
\end{equation*}
To evaluate the term, we note that the SDE given by \eqref{eq:bachelier_model} has the following solution  
\begin{equation*}
    S(T^\mathrm{E})= S(t)+\sigma_\mathrm{B}\left(W^C(T^\mathrm{E})-W^C(t)\right)\coloneqq S(t)+\sigma_\mathrm{B} X,
\end{equation*}
where $X$ follows a normal distribution $N\left(0,T^\mathrm{E}-t \right)$ under $Q^C$ conditional on $\iF_t$. Setting $\omega =1$ and substituting yields 
\begin{align*}
    \EE^C \left[ \left( K-S(T^\mathrm{E})\right)^+   \given \iF_t \right] & = \int_\Omega \left( K-S(t) -\sigma_\mathrm{B}X\right)^+ \dd Q^C \\
    & = \int^{\infty}_{-\infty} \left( K-S(t) -\sigma_\mathrm{B}x\right)^+\frac{1}{\sqrt{2\pi(T^\mathrm{E}-t)}}\exp\left(-\frac{x^2}{2(T^\mathrm{E}-t)}\right)\dd x,
\end{align*}
where the first line follows from the independence of $X$ from $\iF_t$. Applying the substitution $z = -\frac{x}{\sqrt{T^\mathrm{E}-t}}$ leads to
\begin{equation*}
     \EE^C \left[ \left( K-S(T^\mathrm{E})\right)^+   \given \iF_t \right] =  \int^{\infty}_{-\infty} \left(K-S(t)+z\sigma_\mathrm{B}\sqrt{T^\mathrm{E}-t}\right)^+ \frac{1}{\sqrt{2\pi}}\exp\left(-\frac{z^2}{2}\right)\dd z.
\end{equation*}
To further simplify the integral, setting $z_t\coloneqq \frac{S(t)-K}{\sigma_\mathrm{B}\sqrt{T^\mathrm{E}-t}}$ and $\phi(z) = \frac{1}{\sqrt{2\pi}}\exp(\frac{-z^2}{2})$ leads to 
\begin{align*}
     \EE^C \left[ \left( K-S(T^\mathrm{E})\right)^+   \given \iF_t \right] & = \sigma_\mathrm{B}\sqrt{T^\mathrm{E}-t} \int^{\infty}_{-\infty} (-z_t+z)^+\phi(z) \dd z \\
     & = \sigma_\mathrm{B}\sqrt{T^\mathrm{E}-t} \int^{\infty}_{z_t} (-z_t+z)\phi(z)\dd z \\
     & = \sigma_\mathrm{B}\sqrt{T^\mathrm{E}-t} \left[-z_t \int^{\infty}_{z_t}\phi(z)\dd z + \int^{\infty}_{z_t} z\phi(z) \dd z\right] \\ 
     & = \sigma_\mathrm{B}\sqrt{T^\mathrm{E}-t} \left( -z_t\Phi(-z_t)+\phi(z_t)\right).
\end{align*}
\par Finally, performing the analogous calculation for $\omega = -1$ and combining both cases produces
\begin{equation*}
    \ESWO^\mathrm{B}_t (T^\mathrm{E},T^\mathrm{S},\IRS) = C(t)\sigma_\mathrm{B} \sqrt{T^\mathrm{E}-t}\left[  -\omega z_t \Phi(-\omega z_t)+\phi(z_t) \right].
\end{equation*}
\clearpage
\section{Crank--Nicolson method}\label{app:crank_nicolson_method}
\par This appendix provides a detailed derivation of the Crank--Nicolson scheme for solving the following general PDE arising from the valuation of Bermudan swaptions 
\begin{equation*}
    \frac{\partial}{\partial t}u(t,x) + \left[ V(0,t)-a(t)x\right]\frac{\partial}{\partial x}u(t,x)+\frac{\sigma^2(t)}{2}\frac{\partial^2}{\partial x^2} u(t,x)-\left(x+f^\mathrm{M}(0,t)\right)u(t,x) = 0
\end{equation*}
on the domain $(t,x)\in[0,T]\times[x_\mathrm{min},x_\mathrm{max}]$ with a terminal condition $u(T,x) = H(x)$. We closely follow the approach presented in \cite[pp.~43--52]{andersen2010InterestRateModeling} and adapt it to the setting of this paper. To simplify the notation, we isolate the time derivative using a spatial differential operator $\iL u(t,x)$ and switch to the subscript notation for partial derivatives
\begin{align*}
    u_t(t,x) & = -\iL u(t,x) \\
    \iL u(t,x) & \coloneqq  \left[ V(0,t)-a(t)x\right]u_x(t,x)+\frac{\sigma^2(t)}{2} u_{xx}(t,x)-\left(x+f^\mathrm{M}(0,t)\right)u(t,x).
\end{align*}
In what follows, we assume that the computational domain is partitioned by
\begin{align*}
    t_i & = i\Delta_t \quad\text{for}\quad i = 0, \ldots, n \\
    x_j & = x_\mathrm{min} + j\Delta_x \quad \text{for}\quad j = 0, \ldots, m+1
\end{align*}
with $\Delta_t = T/n$ and $\Delta_x=(x_{\mathrm{max}}-x_{\mathrm{min}})/(m+1)$.
\par To approximate the derivatives with respect to the spatial variable $x$, we use central differences at interior grid points, i.e. for $j=1,\ldots, m$ 
\begin{align*}
    u_x(t,x_j) & \approx \frac{u(t,x_{j+1})-u(t,x_{j-1})}{2\Delta_x} \\
    u_{xx}(t,x_j) & \approx \frac{ u(t,x_{j+1}) -2u(t,x_j) +u(t,x_{j-1})}{\Delta_x^2}.
\end{align*}
Substituting these approximations into the spatial operator $\iL u(t,x)$ and rearranging the terms yields
\begin{align*}
    \iL u(t,x_j)
    &\approx \left[ V(0,t)-a(t)x_j\right]
       \frac{u(t,x_{j+1})-u(t,x_{j-1})}{2\Delta_x} \\
    &\quad + \frac{\sigma^2(t)}{2}
       \frac{u(t,x_{j+1})-2u(t,x_j)+u(t,x_{j-1})}{\Delta_x^2} \\
    &\quad - \left(x_j+f^\mathrm{M}(0,t)\right)u(t,x_j) \\
    &= u(t,x_{j-1})
       \left[
           \frac{\sigma^2(t)}{2\Delta_x^2}
           -\frac{V(0,t)-a(t)x_j}{2\Delta_x}
       \right] \\
    &\quad + u(t,x_j)
       \left[
           -\frac{\sigma^2(t)}{\Delta_x^2}
           -\left(x_j+f^\mathrm{M}(0,t)\right)
       \right] \\
    &\quad + u(t,x_{j+1})
       \left[
           \frac{V(0,t)-a(t)x_j}{2\Delta_x}
           +\frac{\sigma^2(t)}{2\Delta_x^2}
       \right] \\
    &\eqqcolon u(t,x_{j-1})d_j(t)
       +u(t,x_j)m_j(t)
       +u(t,x_{j+1})u_j(t).
\end{align*}
\par Special treatment is necessary at the boundary grid points $x_0$ and $x_{m+1}$. We impose zero second derivatives, i.e. $u_{xx}(t,x_0)=u_{xx}(t,x_{m+1})=0$, and apply forward and backward differences, respectively, to approximate the first derivatives
\begin{equation*}
    u_x(t,x_0) \approx \frac{u(t,x_1)-u(t,x_0)}{\Delta_x} \quad\text{and}\quad u_x(t,x_{m+1} )\approx \frac{u(t,x_{m+1})-u(t,x_m)}{\Delta_x}.
\end{equation*}
Again, substituting into $\iL u(t, x)$ leads to 
\begin{align*}
    \iL u(t,x_0)
    &\approx \left[V(0,t)-a(t)x_0\right]
       \frac{u(t,x_1)-u(t,x_0)}{\Delta_x} \\
    &\quad - \left[x_0+f^\mathrm{M}(0,t)\right]u(t,x_0) \\
    &= u(t,x_0)
       \left[
           -\left(x_0+f^\mathrm{M}(0,t)\right)
           -\frac{V(0,t)-a(t)x_0}{\Delta_x}
       \right] \\
    &\quad + u(t,x_1)
       \frac{V(0,t)-a(t)x_0}{\Delta_x} \\
    &\eqqcolon u(t,x_0)m_0(t)
       +u(t,x_1)u_0(t)
\end{align*}
and 
\begin{align*}
    \iL u(t,x_{m+1})
    &\approx \left[V(0,t)-a(t)x_{m+1}\right]
       \frac{u(t,x_{m+1})-u(t,x_m)}{\Delta_x} \\
    &\quad - \left[x_{m+1}+f^\mathrm{M}(0,t)\right]
       u(t,x_{m+1}) \\
    &= -u(t,x_m)
       \frac{V(0,t)-a(t)x_{m+1}}{\Delta_x} \\
    &\quad + u(t,x_{m+1})
       \left[
           \frac{V(0,t)-a(t)x_{m+1}}{\Delta_x}
           -\left(x_{m+1}+f^\mathrm{M}(0,t)\right)
       \right] \\
    &\eqqcolon u(t,x_m)d_{m+1}(t)
       +u(t,x_{m+1})m_{m+1}(t).
\end{align*}
\par Using the preceding discretizations, the spatial differential operator evaluated at the grid points $\left(\iL u(t, x_j)\right)_{j=0}^{m+1}$ admits the following discrete matrix representation 
\begin{equation*}
    \iL(t)\coloneqq \left(\iL u(t, x_j)\right)_{j=0}^{m+1} \approx L(t)U(t)
\end{equation*}
with
\begin{equation*}
    L(t) \coloneqq \begin{pmatrix}
m_0(t)    & u_0(t)    & 0         & \dots    & \dots      & 0           \\
d_1(t)    & m_1(t)    & u_1(t)    & \ddots   &            & \vdots      \\
0         & d_2(t)    & \ddots    & \ddots   & \ddots     & \vdots      \\
\vdots    & \ddots    & \ddots    & \ddots   & \ddots     & 0           \\
\vdots    &           & \ddots    & d_m(t)   & m_{m}(t)   & u_{m}(t)    \\
0         & \dots     & \dots     & 0        & d_{m+1}(t) & m_{m+1}(t)  
\end{pmatrix}\quad\text{and}\quad U(t) \coloneqq \begin{pmatrix}
    u(t,x_0) \\
    u(t,x_1) \\
    \vdots \\
    \vdots \\
    u(t,x_m)\\
    u(t,x_{m+1}) 
\end{pmatrix}.
\end{equation*}
The time domain is discretized at the midpoints $(t_i+t_{i-1})/2$ by central differences, i.e. for $i=1,\ldots, n$ 
\begin{equation*}
    u_t\left(\frac{t_i+t_{i-1}}{2},x\right) \approx \frac{u(t_i,x)-u(t_{i-1},x)}{\Delta_t}.
\end{equation*}
Finally, the Crank--Nicolson scheme yields the following system of equations
\begin{equation*}
    \frac{U(t_i)-U(t_{i-1})}{\Delta_t} \approx - \frac{1}{2}\left[ \iL (t_i)+\iL(t_{i-1})\right].
\end{equation*}
Substituting and rearranging the terms results in the following backward time-stepping scheme
\begin{equation*}
    \left[ I -\frac{\Delta_t}{2} L(t_{i-1}) \right]U(t_{i-1}) = \left[ I +\frac{\Delta_t}{2} L(t_i) \right]U(t_i)
\end{equation*}
for each $i=1,\ldots,n$, where $I$ denotes the identity matrix.
\end{document}